\documentclass[preprint]{aastex}

\shorttitle{Oxygen Abundance Measurements of SHIELD Galaxies}
\shortauthors{Haurberg et al.}

\usepackage[parfill]{parskip}    
\usepackage{longtable}
\usepackage{makecell}
\usepackage{epsfig}
\usepackage{graphicx}
\usepackage{amssymb}
\usepackage{epstopdf}
\usepackage{lscape}
\usepackage{amsmath}
\usepackage{afterpage}
\usepackage{dcolumn}
\newcommand{\hi}{\ion{H}{1}}
\newcommand{\hii}{\ion{H}{2}}

\newcommand{\nha}{[\ion{N}{2}]$\,\lambda6583/\mbox{H}\alpha$}

\newcommand{\ohb}{$\mbox{[\ion{O}{3}]}\,\lambda5007/\mbox{H}\beta$}

\newcommand{\ha}{H$\alpha$}
\newcommand{\oiiite}{[\ion{O}{3}]$\,\lambda4363$}
\newcommand{\oiii}{[\ion{O}{3}]$\,\lambda \,\lambda4959, 5007$} 
\newcommand{\rtt}{([\ion{O}{3}]$\,\lambda \,\lambda4959,5007$+[\ion{O}{2}]]$\,\lambda3727$)/{H}$\beta$}
\newcommand{\ott}{[\ion{O}{3}]$\,\lambda \,\lambda4959,5007$/[\ion{O}{2}]$\,\lambda3727$}
\newcommand{\niioii}{[\ion{N}{2}$]\,\lambda6583$/[\ion{O}{2}]$\,\lambda3727$}
\newcommand{\nii}{[\ion{N}{2}]$\,\lambda6583$}
\newcommand{\te}{T$_{e}$}

\begin{document}

\title{Oxygen Abundance Measurements of SHIELD Galaxies}

\author{Nathalie C. Haurberg\altaffilmark{1,2,4}, John J. Salzer\altaffilmark{1,4}, John M. Cannon\altaffilmark{3,4}, \& Melissa V. Marshall\altaffilmark{3}}

\altaffiltext{1}{Department of Astronomy, Indiana University, 727 E. Third St., Bloomington, IN 47405; nhaurber@astro.indiana.edu, slaz@astro.indiana.edu}
\altaffiltext{2}{Physics Department, Knox College, 2 E. South St., Galesburg, IL 61401; nhaurber@knox.edu}
\altaffiltext{3}{Department of Physics \& Astronomy, Macalester College, 1600 Grand Avenue, Saint Paul, MN 55105; jcannon@macalester.edu}
\altaffiltext{4}{Visiting Astronomer, Kitt Peak National Observatory, National Optical Astronomy Observatory, which is operated by the Association of Universities for Research in Astronomy (AURA) under cooperative agreement with the National Science Foundation.}
                                   
\pagebreak

\begin{abstract}

We have derived oxygen abundances for 8 galaxies from the Survey for \hi\ in extremely low-mass dwarfs (SHIELD). The SHIELD survey is an ongoing study of very low-mass galaxies, with $M_{HI}$ between 10$^{6.5}$ and 10$^{7.5}$ $M_{\odot}$, that were detected by the Arecibo Legacy Fast ALFA (ALFALFA) survey. \ha\ images from the WIYN 3.5m telescope show that these 8 SHIELD galaxies each possess one or two active star-forming regions which were targeted with long-slit spectral observations using the Mayall 4m telescope at KPNO.  We obtained a direct measurement of the electron temperature by detection of the weak \oiiite\ line in 2 of the \hii\ regions. Oxygen abundances for the other \hii\ regions were estimated using a strong-line method.  When the SHIELD galaxies are plotted on a $B$-band luminosity-metallicity diagram they appear to suggest a slightly shallower slope to the relationship than normally seen. However, that offset is systematically reduced when the near-infrared luminosity is used instead.  This indicates a different mass-to-light ratio for the galaxies in this sample and we suggest this may be indicative of differing star-formation histories in the lowest luminosity and surface brightness dwarf irregulars.
\end{abstract}


\clearpage


\section{INTRODUCTION}

Chemical abundance studies of the interstellar medium in galaxies allow a glimpse into the star-formation and chemical enrichment history of these galaxies. The chemical evolution of local-universe galaxies with very low-metallicity gas is particularly important as these galaxies may be objects of cosmological significance. These metal-poor systems represent potential local-universe analogs to galaxies in the early universe and can provide a better understanding of the processes of star-formation and gas-enrichment in the early universe.  Additionally, in order for very metal-poor galaxies to exist in the local universe, processes of either enriched gas outflow or pristine gas inflow are usually invoked, thus these galaxies can provide important constraints to understanding the role of gas inflows and/or outflows on galaxian evolution.

Many authors have searched for very metal-poor or ``extremely metal-deficient" (XMD; 12 + log(O/H) $\le$ 7.65) galaxies in the local universe \citep[e.g.,][]{knia03, ugry03, mel04, lee04, izo06, brown08, pap08, gus11, berg12, izo12}.  Some authors have extended the search out to intermediate redshift \citep{kak07, hu09, xia12}.  Most of these studies have used emission-line surveys, seeking metal-poor galaxies from large samples of strong emission-line sources.  While these searches have led to a significant increase in the number of known metal-poor galaxies, there have been relatively few discovered in the range qualifying as XMD. The seeming dearth of XMD galaxies has raised the question of whether they are intrinsically rare objects in the local universe or if they exist in large numbers but are mostly missed because emission-line surveys are biased toward systems with bright, high surface brightness \hii\ regions (i.e., systems that have recently undergone a very strong starburst) and are relatively insensitive to isolated low-mass, relatively quiescent star-forming galaxies. 

A promising alternative method for finding XMD galaxies has been to study very low-luminosity dwarfs since there is an historically well-established correlation between metallicity and luminosity among low-redshift dwarf galaxies \citep{Leq79, skill89a, pil01, ms02, trem04, salz05c4, lee06c4, vzh06}. This method has already led to the discovery of several XMD galaxies \citep{skill89a, skill89b, vz00, pus05, pus11, berg12, leop_skill}. Despite the fact that there should be large numbers of low-luminosity galaxies according to the galaxy luminosity function, searching for these systems is very difficult as they tend to be very low surface brightness and difficult to detect optically.  However, blind \hi\ surveys do not suffer optical selection biases and thus can uncover low-mass \hi\ sources that potentially have low-surface brightness optical counterparts that may be missed in optical studies.  

It is particularly important to identify and study \emph{low surface brightness} XMD sources because our current understanding of XMD galaxies is derived almost exclusively from blue compact dwarf galaxies (BCDs).  However, normal low-luminosity star-forming galaxies should outnumber BCDs by at least an order of magnitude and thus should be more representative of the XMD population. The Arecibo Dual Beam Survey \citep[ADBS;][]{rs2000} detected many low-mass sources with low-surface brightness dwarf irregular optical counterparts.  Twelve of these ADBS galaxies were selected for spectroscopic follow-up by \citet{hau13} to determine their chemical abundances. While none of these galaxies were found to be XMD galaxies, the sample was generally low-metallicity and showed that \hi\,-\,surveys can efficiently uncover metal-poor and potential XMD galaxies.  The more recent Arecibo Legacy Fast ALFA (ALFALFA) blind \hi\,-\,survey \citep{giov05, haynes11} has uncovered hundreds of low-mass \hi\ sources, many with very low-surface brightness optical counterparts. The ALFALFA catalog provides a promising target list of potential XMD galaxies.  

The use of \hi\ catalogs to search for low-luminosity XMD systems has already proven successful with the recent discovery of the Leo P dwarf. Leo P was selected as a candidate XMD from the ALFALFA catalog due to its low derived \hi mass \citep{leop_giov} and very-low stellar mass inferred from optical follow-up imaging \citep{leop_rhode}.  Optical spectroscopy revealed an extremely low oxygen abundance of 12 + log(O/H) = 7.17$\,\pm\,$0.04 \citep{leop_skill} confirming it as an XMD galaxy and one of the most metal-poor galaxies known.  
 
 In this paper, we report the findings from a program dedicated to determining and analyzing the nebular abundances of a set of low-luminosity, low surface-brightness, low \hi\ mass galaxies selected from the ALFALFA catalog as part of The Survey for \hi\ in Extremely Low-Mass Dwarfs (SHIELD). SHIELD is a comprehensive multi-wavelength project targeting 12 of the lowest-\hi\ mass galaxies discovered in the ALFALFA survey which have a clear optical counterpart \citep{cannon11}.  The optical and \hi\ properties of these galaxies (N.~Haurberg et al.~in preparation) make them prime candidates as XMDs.

In Section \ref{sec:sample} we describe the sample selection in more detail and include the derived photometric and \hi\ properties. The spectroscopic observations are described in Section \ref{sec:obs} and the reduction and measurement processes are outlined in Section \ref{sec:reduct}.  Section \ref{sec:abun} describes how we determined abundances for the \hii\ regions and the abundance results.  The analysis and discussion of the results are in Section \ref{sec:analysis}, and our conclusions are in Section \ref{sec:concl}.

\section{SAMPLE SELECTION}
\label{sec:sample}

The ALFALFA survey includes hundreds low-\hi-mass objects with $M_{HI} < 10^{8} M_{\odot}$ and additionally has provided the first robust sample of galaxies at the \textit{very} low-mass end of the \hi\ mass function with $M_{HI} < 10^{7} M_{\odot}$ \citep{mar10}. Twelve of these galaxies, with 10$^{6.5} M_{\odot} < M_{HI} < 10^{7.5} M_{\odot}$, were selected for the SHIELD project \citep{cannon11} from a preliminary ALFALFA catalog.  The galaxies selected for SHIELD were those with the lowest \hi\ mass that had apparent optical counterparts in Sloan Digitized Sky Survey (SDSS) images.  These galaxies represent the lowest mass potentially star-forming galaxies that have formed a significant number of stars. Thus, following the traditional luminosity-metallicity trend (or mass-metallicity trend), the galaxies in this sample should be some of the most metal-poor star-forming galaxies in the local universe.

The SHIELD project encompasses a complex multi-wavelength data set including detailed \hi\ gas mapping from multi-configuration Expanded Very Large Array\footnotemark[5] observations \citep{cannon11}, broadband BVR and narrowband H$\alpha$ imaging from the WIYN 3.5m telescope at Kitt Peak National Observatory\footnotemark[6] (KPNO; N.~Haurberg et al.~in preparation), 3.6 and 4.5$\micron$ imaging with the InfraRed Array Camera (IRAC) on the Spitzer Space Telescope\footnotemark[7], Hubble Space Telescope\footnotemark[8] Imaging with the Wide Field Camera 3 using the F606W ($\sim\,R$) and F814W ($\sim\,I$) filters, and long-slit optical spectra with the Richey-Chretien Focus Spectrograph on the Mayall 4m at KPNO (this work).  
\footnotetext[5]{The National Radio Astronomy Observatory is a facility of the National Science Foundation operated under cooperative agreement by Associated Universities, Inc.}
\footnotetext[6]{The WIYN Observatory is a joint facility of the University of Wisconsin-Madison, Indiana University, Yale University, and the National Optical Astronomy Observatory.}
\footnotetext[7]{This work is based in part on observations made with the Spitzer Space Telescope, which is operated by the Jet Propulsion Laboratory, California Institute of Technology under a contract with NASA. Support for this work was provided by NASA.}
\footnotetext[8]{Based on observations made with the NASA/ESA Hubble Space Telescope, obtained at the Space Telescope Science Institute, which is operated by the Association of Universities for Research in Astronomy, Inc., under NASA contract NAS 5-26555. These observations are associated with program 12658.}


Since these galaxies were selected from a blind \hi\ survey, the sample does not suffer any bias toward high-surface brightness objects as is often present in optical catalogs.  The only optical qualification for galaxies in the SHIELD sample is that there be some evidence of a stellar counterpart associated with the \hi\ source in SDSS images.  Low \hi\ mass sources in the ALFLFA catalog with no apparent optical counterpart have been studied by \citet{adams13} and are also of great interest as these may represent sources with extremely low surface brightness counterparts or starless dark matter halos.  

\begin{figure}
\centering
\includegraphics[width=4.2in]{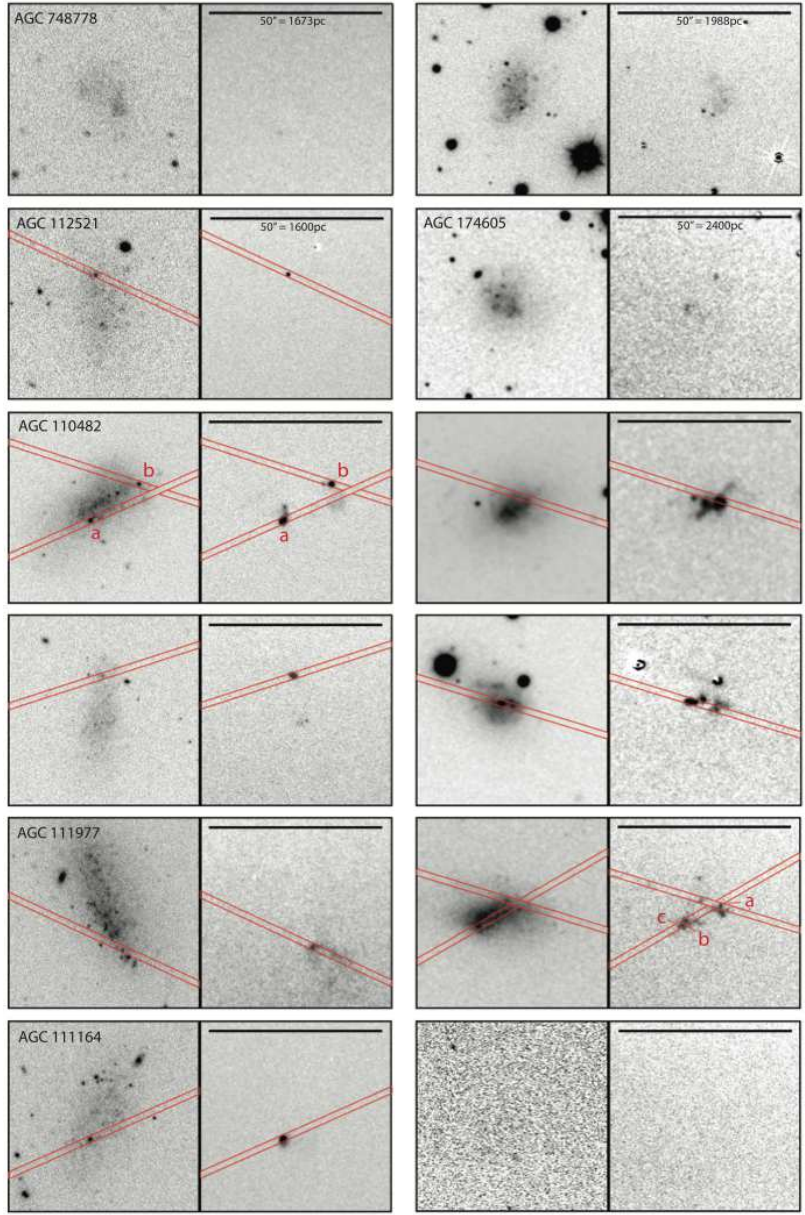}
\caption[Thumbnail \ha\ Images of the SHIELD galaxies]{Thumbnail images of the SHIELD galaxies, oriented so that North points up and East points to the left. On the left of each column are short exposure $R$-band images and on the right are continuum subtracted H$\alpha$ images from the WIYN 3.5m (N.~Haurberg et al.~in preparation). Each thumbnail is approximately 50$\arcsec \times 50\arcsec$. The approximate position of the slit is indicated for the 8 SHIELD galaxies for which we obtained usable spectra. Those objects with no slit image superimposed either had no visible H$\alpha$ emission or did not produce usable spectra.}
\label{fig:finder}
\end{figure}

Thumbnail images showing the WIYN $R$-band and \ha\ images for the 12 SHIELD galaxies are shown in Figure \ref{fig:finder}.  The \hii\ regions identified from the displayed \ha\ images were targeted for the long-slit spectroscopic follow-up that is the subject of this paper.  Figure \ref{fig:finder} also shows the approximate slit locations overlaid on each galaxy.  A comprehensive set of general and photometric quantities that describe the SHIELD galaxies is compiled in Table \ref{tab:c4basic}. Column 1 lists the galaxies in the SHIELD sample with the coordinates for each source listed in Columns 2 and 3.  The distance, calculated using HST optical imaging to determine the tip of the red giant branch \citep{mcquinn}, is given in Column 4 and is used as the assumed distance for the derivation of all distance-dependent quantities.  The absolute $B$-band magnitude, from the WIYN 3.5m (N.~Haurberg et al.~in preparation) images is listed in Column 5; the listed uncertainty includes the error in both the distance and photometric error. The \bv\ color and $B$-band central surface brightness not corrected for inclination effects ($\mu_{0,B}$) are listed in Columns 6 and 7, also from N.~Haurberg et~al (in preparation).  Columns 8 and 9 contain the 4.5$\micron$ and 3.6$\micron$ flux (in mJy) from the Spitzer Space Telescope observations, and Column 10 lists the stellar mass estimate derived from the Spitzer data (Cannon, Marshall, et al.~in preparation).  The stellar masses presented in this table were calculated using the method of \citet{esk12}.  The \hi\ masses given in Column 10 are from \citet{cannon11}, but have been adjusted for the updated distances. Column 11 lists the star-formation rate (SFR) calculated from the \ha\ images (N.~Haurberg et al.~in preparation) using the standard prescription of  \citet{ken98} to convert from \ha\ luminosity to SFR.

The optical images from the WIYN 3.5m telescope confirmed the optical counterparts of the SHIELD \hi\ sources as low surface brightness dIrrs with blue colors (Table \ref{tab:c4basic}).  The \ha\ imaging revealed that the star formation in these galaxies was generally confined to 1 or 2 knots of star formation which were targeted with long-slit spectroscopic follow-up observations presented in this work.  Two of the SHIELD galaxies, AGC 749241 and AGC 748778 were not included in this spectroscopic follow-up as the \ha\ imaging revealed no star-forming nebulae.  AGC 749241 had no detected \ha\ emission associated with it and AGC 748778 displayed only very weak, diffuse \ha\ emission (2.6 $\sigma$ above the sky) with no discrete \hii\ regions (see Figure \ref{fig:finder}).  Additionally, \hii\ regions were detected in AGC 174585 and AGC 174605 but they were too faint to provide useful spectra so they are not included in the following analysis. Hence, we were able to derive useful abundance estimates for 8 of the 12 SHIELD galaxies.

\afterpage{
\begin{landscape}
\begin{table}[c]
\scriptsize
\begin{longtable}[pc]{l c c c c c c c c c c c c}
\caption{SHIELD Galaxy Properites} \label{tab:c4basic} \\
\hline \hline \noalign{\smallskip}
Galaxy		&	R.A.			& Dec		& Distance$^{1}$    	&	$M_{B}^{2}$		&\bv$^{2}$			&$\mu_{0,B}^{2}$		&	f3.6$^{3}$		&  f4.5$^{3}$	& log($M_{\star})^{3}$ 	&	log($M_{HI})^{4}$   &   log(SFR)$^{2}$	   \\   
		&   	[2000.0]		& [2000.0]	& [Mpc]			& 	[mag]			&[mag]  			& [mag $''^{-2}$] 				& 	[mJy] 		& [mJy]		& [$M_{\odot}$] 		& 	[$M_{\odot}$]	    &    [$M_{\odot}\,\textrm{yr}^{-1}$]  \\
\noalign{\smallskip} \hline \noalign{\smallskip}	
AGC 748778	&    	00:06:34.3		& 15:30:39 & 	6.46$\,\pm\,$0.15		 &	-10.34$\,\pm\,$0.08	  &   0.22$\,\pm\,$0.03	 	&  23.78        &0.0453	                     &	0.0310		      & 5.82           		&	6.72		    &  -4.92$^{\textrm{up. lim.}}$  \\
AGC 112521	& 	01:41:07.6		& 27:19:24 &	6.58$\,\pm\,$0.18		 &	-10.59$\,\pm\,$0.08	  &   0.45$\,\pm\,$0.03	 	&  24.55		&0.563 		  	&	0.402			& 6.93     			&	7.11		    &  -4.10$\,\pm\,$0.08\\
AGC 110482	& 	01:42:17.4		& 26:22:00 &	7.82$\,\pm\,$0.21		 &	-13.02$\,\pm\,$0.13	  &   0.49$\,\pm\,$0.02	 	&  22.45		&1.05 	  	  	&	0.727			& 7.35	       	    	&	7.31		    &  -2.67$\,\pm\,$0.04\\
AGC 111946	& 	01:46:42.2		& 26:48:05 &	9.02$\,\pm\,$0.25		 &	-11.87$\,\pm\,$0.12	  &   0.30$\,\pm\,$0.02	 	&  24.06		&0.235 	  	  	&	0.164			& 6.83	      	    	&	7.11		    &  -3.10$\,\pm\,$0.07\\
AGC 111977	& 	01:55:20.2		& 27:57:14 &	5.96$\,\pm\,$0.10		 &	-12.60$\,\pm\,$0.09	  &   0.48$\,\pm\,$0.02	 	&  22.67		&1.04 	  	  	&	0.669			& 7.11	      	    	& 	6.85		    &  -3.06$\,\pm\,$0.08\\
AGC 111164	& 	02:00:10.1		& 28:49:52 &	5.11$\,\pm\,$0.07		 &	-11.16$\,\pm\,$0.10	  &   0.42$\,\pm\,$0.02	 	&  23.83		&0.470 	  	  	&	0.355			& 6.64	      	    	&	6.64		    &  -3.40$\,\pm\,$0.05\\
AGC 174585	& 	07:36:10.3		& 09:59:11 &	7.89$\,\pm\,$0.19		 &      -11.32$\,\pm\,$0.13     &   0.41$\,\pm\,$0.04 	&  23.17		&0.139 	  	  	&	0.0944			& 6.48             		&	6.94		    &  -3.21$\,\pm\,$0.05\\
AGC 174605	& 	07:50:21.7		& 07:47:40 &	10.89$\,\pm\,$0.28		 &	-12.22$\,\pm\,$0.11	  &   0.39$\,\pm\,$0.03	 	&  22.29		&0.207 	  	  	&	0.147			& 6.94	      	    	&	7.19		    &  -3.19$\,\pm\,$0.04\\
AGC 182595	& 	08:51:12.1		& 27:52:48 &	9.02$\,\pm\,$0.28		 &	-12.70$\,\pm\,$0.13	  &   0.52$\,\pm\,$0.03	 	&  22.31		&0.622 	  	  	&	0.408			& 7.25	      	   	&	6.92		    &  -2.65$\,\pm\,$0.07\\		 
AGC 731457	& 	10:31:55.8		& 28:01:33 &	11.13$\,\pm\,$0.18		 &	-13.73$\,\pm\,$0.11	  &   0.38$\,\pm\,$0.03	 	&  21.31		 	&$\cdots$ 	&	$\cdots$		  	& $\cdots$	      	&	7.09		    &  -2.52$\,\pm\,$0.07\\
AGC 749237	& 	12:26:23.4		& 27:44:44 &	11.62$\,\pm\,$0.18		 &	-14.12$\,\pm\,$0.12	  &   0.45$\,\pm\,$0.03	 	&  21.53	 	&0.837 		  	&	0.544			& 7.60	      	    	&	7.49		    &  -2.34$\,\pm\,$0.05\\
AGC 749241	& 	12:40:01.7		& 26:19:19 &	5.62$\,\pm\,$0.15		 &	-10.25$\,\pm\,$0.19	  &   0.14$\,\pm\,$0.03	 	&  23.37		 	& $\cdots$	&	$\cdots$		  	& $\cdots$	      	&	6.74		    &  -5.00$^{\textrm{up. lim.}}$      \\

\noalign{\smallskip} \hline \hline
\noalign{\smallskip} 
\multicolumn{12}{p{1.23\textwidth}}{$^{1}$\citet{mcquinn} derived from HST images, tip of the red giant branch; $^{2}$N.~Haurberg et al (in preparation) derived from WIYN 3.5m observations; $^{3}$ Cannon, Marshall, et al.~in prep.; derived from Spitzer Space Telescope Images; $^{4}$\citet{cannon11}, derived from radio observations and adjusted for new distance estimate}
\end{longtable}
\end{table}
\end{landscape}
}
\normalsize


\section{OBSERVATIONS}
\label{sec:obs}

\subsection{OPTICAL IMAGING}

Optical imaging observations were performed using the Mini-Mosiac Imager on the WIYN 3.5m telescope at Kitt Peak National Observatory (KPNO) over four nights: 2010 October 7-8 and 29-30 in March 2011.  The two Fall nights (Oct.~2010) had fantastic seeing ($\approx0.5''$) and were done under photometric conditions, while the two nights in the Spring (Mar. 2011) had degraded seeing ($\approx1.2''$) and were not photometric.  Short-exposure post-calibration observations for the Spring targets were performed in April of that 2011.  The broadband Johnson $B$, $V$, $R$, and W036 narrowband \ha\ filters were used. The nominal field of view was 9.6$'\times$ 9.6$'$ across 2 chips and 4 amplifiers.

Exposure times for the broadband images in the fall were 900, 720, and 600 s for $B$, $V$, and $R$ bands, respectively.  The narrowband imaging was taken in sets of two long narrowband \ha\ exposures (900 s) with a short (180 s) R-broadband image between them.  Since the conditions were less ideal in the Spring, longer exposure times were used.  The Spring broadband data were taken with 1200, 720, and 900 s exposures (for $B$, $V$, and $R$, respectively) and 1200 s narrowband \ha\ exposures with a 240 s $R$-band exposure in between.

The broadband observations times should be sufficiently long to have detected any diffuse emission from the galaxies. These optical broadband images were analyzed with both isophotal fitting techniques as well as large aperture photometry as discussed in detail in N.~Haurberg et al.~in preparation.  While we can not rule out that there may be an extraordinarily low surface brightness component that was not detected in our images, comparison of with the subsequent HST imaging (?????) shows now indication of such, thus we remain confident our images were sufficient at detecting all of the optical emission in these galaxies.

\subsection{SPECTRAL OBSERVATIONS}

The spectral observations presented in this work were carried out using the Mayall 4m telescope at Kitt Peak National Observatory with the Richey-Chretien Focus Spectrograph and T2KA imager over the course of 5 nights: 2012 April 17\,--\,19 and 2012 October 15\,--\,16. The KPC-10A grating (316 lines mm$^{-1}$) and WG-345 blocking filter were used. The grating is blazed at 4000 \AA\ giving a dispersion of 2.78 \AA\ pixel$^{-1}$ and total coverage from 2850\,--\,8550 \AA\ on the CCD.  All spectra were taken with a slit-width of 1.5\arcsec\ and the slit extended 324\arcsec\ along the spatial direction. 

Target sources were too dim to be seen with the acquisition cameras, so nearby bright stars were used for blind-offsets. Offsets were checked by using multiple stars in the field, and were successful in aligning sources within the slit.  The slit was positioned as nearly along the parallactic angle as possible to avoid the effects of differential refraction through the atmosphere. Since the star-forming regions in these galaxies are relatively isolated, usually only the target source fell in the slit, but in two cases an additional emission nebula ended up in the slit as well. 

We observed several spectrophotometric standard stars throughout each night, which were used to calibrate the flux scale for our spectra.  The standards were selected from the lists of \citet{og83} and \citet{mass88}.  Each night we additionally took images of HeNeAr spectral lamps for wavelength calibration, zero-length exposure bias images, internal quartz lamps for flat field calibration, and twilight sky images to correct for slit-width variations.

\section{SPECTRAL REDUCTION AND MEASUREMENT}
\label{sec:reduct}

Spectral images were processed through the standard spectral reduction routines in \textsc{iraf}\footnotemark[9].  The bias level was determined from the overscan region in each image and the mean bias image was used to account for any two-dimensional (2D) structure in the bias. Median combined quartz lamp flats were used to account for pixel-to-pixel variations and twilight sky flats were used to create an illumination function correcting for variations in the slit width.  The science images were then processed through the \textsc{lacos\_spec} cosmic ray rejection routine of \citet{vd01} and the resultant ``cleaned" images were examined carefully by eye in comparison with the original image to ensure that emission line pixels were not rejected as cosmic rays.
\footnotetext[9]{\textsc{iraf} is distributed by the National Optical Astronomy Observatories, which are operated by the Association of Universities for Research in Astronomy, Inc., under cooperative agreement with the National Science Foundation.}

\hii\ region spectra were extracted using the \textsc{apall} package in \textsc{iraf}.  Since the diffuse background emission was typically very weak, local sky regions that closely bracketed the emission line source were selected for night-sky subtraction. The wavelength scale was determined for each night using the solution derived from HeNeAr lamp spectra. The spectrophotometric standard stars were then used to create a sensitivity function that was applied to all the spectra.  After the spectra were wavelength and flux calibrated, standard stars were compared to a catalog spectrum in order to check the calibration, which appeared good in all cases.  As a last step, telluric absorption features were removed.

\afterpage{
\begin{figure}
\centering
\includegraphics[width=6in]{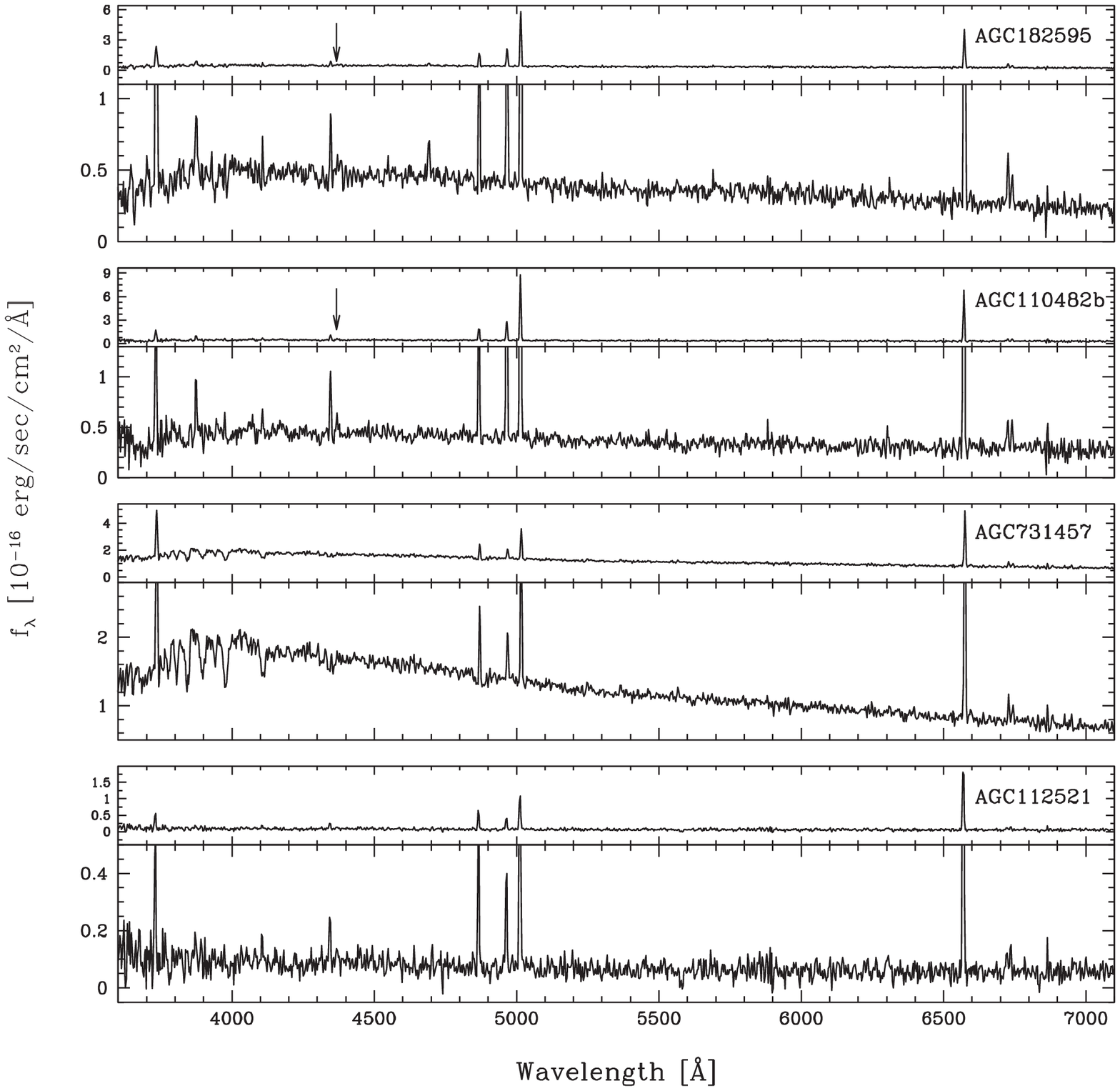}
\caption[Example spectra of SHIELD galaxies]{Four examples of spectra from \hii\ regions in the SHIELD galaxies. The first two display a measurable [\ion{O}{3}]$\lambda$4363 emission line and the location of this line is marked with a small arrow.}
\label{fig:spec}
\end{figure}
}

Figure \ref{fig:spec} shows examples of the final spectra. Several important strong lines were present in all the extracted spectra, most notably \ha, H$\beta$, the [\ion{O}{2}] doublet, and the strong lines from the [\ion{O}{3}] triplet. The weak, temperature sensitive, \oiiite\ line was discernible in two of the \hii\ regions, AGC 110482\emph{b} and AGC 182595.  The other extracted spectra were good quality but did not contain an unambiguous detection of the \oiiite\ line.  In total, we extracted 11 unique \hii\ regions from 8 galaxies.  Both of the \hii\ regions with discernible \oiiite\ are included in Figure \ref{fig:spec}. We have plotted a version showing the full dynamic range of the emission and another version that displays only a small range in flux.

Once emission lines were identified they were measured using the \textsc{splot} routine in \textsc{iraf}.  This routine allows interactive setting of the continuum level at each line.  To decide on the correct location for the continuum we used an average of the local continuum on either side of the emission line and assumed a linear fit for for the continuum under the emission line.  We generally assumed that the continuum remained at a constant level under the emission line except for those cases where the lines were very broad or the slope of the continuum was steep enough to change significantly through the line. 

Since we are measuring emission from regions of relatively recent star-formation it is expected that the permitted Balmer lines are affected by underlying absorption from hot main-sequence stars.  The underlying absorption was accounted for by first assuming that the equivalent width of absorption is the same for all the Balmer lines and then using multiple Balmer emission-line ratios (H$\alpha$/H$\beta$, H$\gamma$/H$\beta$, and H$\delta$/H$\beta$)  to simultaneously estimate the amount of underlying absorption and c(H$\beta$) (i.e. the absorption coefficient at H$\beta$).  We were able to estimate the proper absorption correction by determining where the c(H$\beta$) coefficients calculated from each Balmer ratio converged.  This correction was applied to the hydrogen lines so that the relative Balmer intensities could be used to accurately determine the reddening for each \hii\ region. In cases where only one Balmer ratio was available, an average absorption correction of 1.0 \AA\ was used since most \hii\ regions showed some absorption that we felt needed to be accounted for.  The reddening was determined using the reddening law of \citet{car89} and using the temperature derived from the  [\ion{O}{3}] lines when possible. If the \oiiite\ line was not detected, we assumed an electron temperature of 10,000 K.  

Since the Balmer decrement was used to calculate the reddening, there is no distinction between intrinsic and foreground reddening.  The reddening corrected flux measurements (relative to H$\beta$) for each line we measured are tabulated for each \hii\ region in Table \ref{tab:c4gold}, along with the assumed equivalent width of underlying absorption (W$_{\textrm{abs}}$),  c(H$\beta$), and the assumed \te\ and n$_{e}$.  The electron temperature ($T_{e}$) and electron density (n$_{e}$) of the nebula were estimated using the \oiiite$/$\oiii\ and [\ion{S}{2}]$\,\lambda \,\lambda6716, 6731$ ratios, respectively. For cases where the \oiiite\ line was measured, the temperature and density calculation were performed with the ELSA program \citep{elsa} described more completely in the following section.  In cases where there was not a reliable measurement of [\ion{S}{2}]$\,\lambda \,\lambda6716, 6731$ a default electron density of 100 cm$^{-3}$ was assumed.  This density is consistent with what is commonly observed for \hii\ regions.  The errors given on the line fluxes were calculated by fully propagating the derived errors from the relevant quantities (e.g., rms of the continuum, rms scatter in flux calibration, etc.).

\afterpage{
\scriptsize
\begin{landscape}
\begin{singlespace}
\begin{longtable}{l c c c c c c}
\caption{Line Measurement Results} \label{tab:c4gold} \\

\hline \hline \noalign{\smallskip}
Line Identification						&  AGC 112521	                   	  &   AGC 110482\emph{a}		  &   AGC 110482\emph{b}	  &   AGC 111946 				&   AGC1 11977                 		 &      AGC 111164               			\\	       
(\AA) 								& {F$_{\textrm{ion}}$/H$\beta$}  & {F$_{\textrm{ion}}$/H$\beta$} 	& {F$_{\textrm{ion}}$/H$\beta$}  & {F$_{\textrm{ion}}$/H$\beta$} 	& {F$_{\textrm{ion}}$/H$\beta$} & {F$_{\textrm{ion}}$/H$\beta$} 		\\
\noalign{\smallskip} \hline \noalign{\smallskip}         
$[$\ion{O}{2}$]\,\lambda 3728$ 			&  0.809 $\,\pm\,$ 0.075  		&   3.062 $\,\pm\,$ 0.217			&   1.306 $\,\pm\,$ 0.091     	&   3.137 $\,\pm\,$ 0.282  			&   1.965 $\,\pm\,$ 0.250      	&      1.417 $\,\pm\,$ 0.095	 			\\
$[$\ion{Ne}{3}$]\,\lambda 3869$ 			&  $\cdots$	    	  		&   0.300 $\,\pm\,$ 0.031			&   0.555 $\,\pm\,$ 0.044     	&   $\cdots$	     	   			&   1.161 $\,\pm\,$ 0.150      	&      0.213 $\,\pm\,$ 0.019	 			\\
\ion{He}{1}+H$\zeta$	  				&  $\cdots$	    	   		&   $\cdots$     	    				&   $\cdots$	       	      		&   $\cdots$	     	   			&   $\cdots$		       	  	&      0.152 $\,\pm\,$ 0.015	 			\\
$[$\ion{Ne}{3}$]\,\lambda 3970$+H$\epsilon$	&  $\cdots$	    	  		&   $\cdots$	     				&   0.077 $\,\pm\,$ 0.023     	&   $\cdots$	     	   			&   $\cdots$		       	  	&      0.133 $\,\pm\,$ 0.013	 			\\
H$\delta$  		  					&  0.260 $\,\pm\,$ 0.027  		&   0.432 $\,\pm\,$ 0.034			&   0.252 $\,\pm\,$ 0.027     	&   0.298 $\,\pm\,$ 0.045  			&   $\cdots$		       	  	&      0.249 $\,\pm\,$ 0.018	 			\\
H$\gamma$ 							&  0.387 $\,\pm\,$ 0.032  		&   0.433 $\,\pm\,$ 0.031			&   0.520 $\,\pm\,$ 0.034     	&   0.508 $\,\pm\,$ 0.052  			&   $\cdots$		       	  	&      0.436 $\,\pm\,$ 0.025	 			\\
$[$\ion{O}{3}$]\,\lambda 4363$ 			&  $\cdots$	    	  		&   $\cdots$     					&   0.109 $\,\pm\,$ 0.021		&   $\cdots$    	     	   			&   $\cdots$	     	       	  	&      $\cdots$ 		 				\\
H$\beta$    	      						&  1.000 $\,\pm\,$ 0.060  		&   1.000 $\,\pm\,$ 0.051			&   1.000 $\,\pm\,$ 0.045     	&   1.000 $\,\pm\,$ 0.068  			&   1.000 $\,\pm\,$ 0.164      	&      1.000 $\,\pm\,$ 0.044	 			\\
$[$\ion{O}{3}$]\,\lambda 4959$ 			&  0.527 $\,\pm\,$ 0.036  		&   0.939 $\,\pm\,$ 0.049			&   1.640 $\,\pm\,$ 0.066     	&   0.276 $\,\pm\,$ 0.034  			&   0.857 $\,\pm\,$ 0.090      	&      0.750 $\,\pm\,$ 0.035	 			\\
$[$\ion{O}{3}$]\,\lambda 5007$ 			&  1.701 $\,\pm\,$ 0.092  		&   2.708 $\,\pm\,$ 0.124			&   5.189 $\,\pm\,$ 0.189     	&   0.838 $\,\pm\,$ 0.061  			&   3.375 $\,\pm\,$ 0.270      	&      2.339 $\,\pm\,$ 0.096	 			\\
\ion{He}{1}$\,\lambda 5876$      			&  0.113 $\,\pm\,$ 0.017  		&   $\cdots$	    				&   0.051 $\,\pm\,$ 0.014     	&   $\cdots$	     	   			&   $\cdots$		       	  	&      $\cdots$	 		 			\\
H$\alpha$  							&  2.860 $\,\pm\,$ 0.225  		&   2.860 $\,\pm\,$ 0.201			&   2.785 $\,\pm\,$ 0.169     	&   2.860 $\,\pm\,$ 0.237  			&   2.860 $\,\pm\,$ 0.510      	&      2.860 $\,\pm\,$ 0.183   			\\
$[$\ion{N}{2}$]\,\lambda 6583$  			&  0.015 $\,\pm\,$ 0.012  		&   0.080 $\,\pm\,$ 0.014			&   $\cdots$	       	      		&   0.226 $\,\pm\,$ 0.029  			&   0.176 $\,\pm\,$ 0.044   	  	&      0.062 $\,\pm\,$ 0.008	 			\\
$[$\ion{S}{2}$]\,\lambda 6717$				&  0.143 $\,\pm\,$ 0.020  		&   0.249 $\,\pm\,$ 0.024			&   0.157 $\,\pm\,$ 0.016     	&   0.227 $\,\pm\,$ 0.029  			&   $\cdots$		       	  	&      0.128 $\,\pm\,$ 0.012	 			\\
$[$\ion{S}{2}$]\,\lambda 6731$   			&  0.131 $\,\pm\,$ 0.019  		&   0.179 $\,\pm\,$ 0.019			&   0.124 $\,\pm\,$ 0.015     	&   0.153 $\,\pm\,$ 0.024  			&   $\cdots$		       	  	&      0.094 $\,\pm\,$ 0.010	 			\\
$[$\ion{Ar}{3}$]\,\lambda 7136$			&  $\cdots$	    	  		&   $\cdots$					&   0.060 $\,\pm\,$ 0.012     	&   $\cdots$           	   			&   $\cdots$		      	  	&      $\cdots$    		 				\\
\cline{2-7}\noalign{\smallskip} 
\te 									& 10,000 K$^{*}$ 			&  10,000 K$^{*}$				& 15,630 K				& 10,000 K$^{*}$ 				& 10,000 K$^{*}$ 			& 10,000 K$^{*}$ \\
n$_{e}$ 								& 389 					& 48							& 146	 				& 100$^{**}$ 					& 100$^{**}$  				& 71  \\
W$_{\textrm{abs}}$\,(\AA) 					& 5.0 					& 2.5							& 1.0 					& 0.0							& 1.0						& 2.5 \\
c(H$\beta$) 							&  0.069  					& 0.277  						& 0.496 					& 0.532						& 0.375		 			& 0.171  \\
\noalign{\smallskip} \hline 
\multicolumn{7}{r}{{Continued on next page}} \\
\newpage

\multicolumn{6}{c}{{-- Continued from previous page}} \\
\cline{1-6}\noalign{\smallskip} \cline{1-6} \noalign{\smallskip}
Line Identification						&   AGC 182595             		 & AGC 731457				& 	AGC 749237\emph{a}	   &  AGC 749237\emph{b}	  & 	AGC 749237\emph{c}      \\	     	     		     	       
(\AA) 								& {F$_{\textrm{ion}}$/H$\beta$} & {F$_{\textrm{ion}}$/H$\beta$} 	& {F$_{\textrm{ion}}$/H$\beta$}   & {F$_{\textrm{ion}}$/H$\beta$} & {F$_{\textrm{ion}}$/H$\beta$} \\
\noalign{\smallskip} \cline{1-6} \noalign{\smallskip}     	     		     	         
$[$\ion{O}{2}$]\,\lambda 3728$ 			&   1.941 $\,\pm\,$ 0.183  		& 3.705 $\,\pm\,$ 0.353			& 	3.283 $\,\pm\,$ 0.365	&  3.326 $\,\pm\,$ 0.314		& 	3.212 $\,\pm\,$ 0.354	\\
$[$\ion{Ne}{3}$]\,\lambda 3869$ 			&   0.493 $\,\pm\,$ 0.048  		& $\cdots$					& 	$\cdots$      	  	  	&  0.622 $\,\pm\,$ 0.060		&  	0.475 $\,\pm\,$ 0.062	\\
\ion{He}{1}+H$\zeta$	  				&   $\cdots$	       	     		& $\cdots$					& 	$\cdots$		  	  	&  $\cdots$     				& 	$\cdots$	  			\\
$[$\ion{Ne}{3}$]\,\lambda 3970$+H$\epsilon$	&   $\cdots$	       	     		& $\cdots$					& 	$\cdots$		  	  	&  $\cdots$				& 	$\cdots$				\\
H$\delta$  		  					&   0.256 $\,\pm\,$ 0.026  		& 0.138 $\,\pm\,$ 0.028			& 	$\cdots$		  	  	&  0.226 $\,\pm\,$ 0.026		& 	$\cdots$				\\
H$\gamma$ 							&   0.429 $\,\pm\,$ 0.036  		& 0.319 $\,\pm\,$ 0.036			& 	0.421 $\,\pm\,$ 0.051	&  0.435 $\,\pm\,$ 0.038		& 	$\cdots$				\\
$[$\ion{O}{3}$]\,\lambda 4363$ 			&   0.087 $\,\pm\,$ 0.015  		& $\cdots$					&     	$\cdots$	  			&  $\cdots$ 				& 	$\cdots$				\\
H$\beta$    	      						&   1.000 $\,\pm\,$ 0.067  		& 1.000 $\,\pm\,$ 0.069			& 	1.000 $\,\pm\,$ 0.085	&  1.000 $\,\pm\,$ 0.070		& 	1.000 $\,\pm\,$ 0.086	\\
$[$\ion{O}{3}$]\,\lambda 4959$ 			&   1.373 $\,\pm\,$ 0.091  		& 0.593 $\,\pm\,$ 0.047			& 	0.892 $\,\pm\,$ 0.079	&  0.796 $\,\pm\,$ 0.057		& 	0.709 $\,\pm\,$ 0.065	\\
$[$\ion{O}{3}$]\,\lambda 5007$ 			&   4.159 $\,\pm\,$ 0.264  		& 1.784 $\,\pm\,$ 0.120			& 	2.487 $\,\pm\,$ 0.194	&  2.102 $\,\pm\,$ 0.139		& 	2.118 $\,\pm\,$ 0.165	\\
\ion{He}{1}$\,\lambda 5876$      			&   $\cdots$	       	     		& $\cdots$					& 	$\cdots$		  	  	&  $\cdots$	 			& 	$\cdots$				\\
H$\alpha$  							&   2.786 $\,\pm\,$ 0.264  		& 2.860 $\,\pm\,$ 0.270			& 	2.860 $\,\pm\,$ 0.315	&  2.860 $\,\pm\,$ 0.277		& 	2.860 $\,\pm\,$ 0.319	\\
$[$\ion{N}{2}$]\,\lambda 6583$  			&   0.064 $\,\pm\,$ 0.013  		& 0.143 $\,\pm\,$ 0.023			& 	0.197 $\,\pm\,$ 0.036	&  0.128 $\,\pm\,$ 0.020		& 	0.120 $\,\pm\,$ 0.029	\\
$[$\ion{S}{2}$]\,\lambda 6717$				&   0.252 $\,\pm\,$ 0.029  		& 0.210 $\,\pm\,$ 0.029			& 	0.186 $\,\pm\,$ 0.036	&  0.446 $\,\pm\,$ 0.050		& 	0.313 $\,\pm\,$ 0.047	\\
$[$\ion{S}{2}$]\,\lambda 6731$   			&   0.193 $\,\pm\,$ 0.024  		& 0.160 $\,\pm\,$ 0.025			& 	0.179 $\,\pm\,$ 0.035	&  0.319 $\,\pm\,$ 0.037		& 	0.132 $\,\pm\,$ 0.030	\\
$[$\ion{Ar}{3}$]\,\lambda 7136$			&   0.085 $\,\pm\,$ 0.015  		& $\cdots$					& 	$\cdots$		  	  	&  $\cdots$ 				& 	$\cdots$				\\
\cline{2-6}\noalign{\smallskip} 
\te 									& 15,680 K 				& 10,000 K$^{*}$ 				& 10,000 K$^{*}$ 			& 10,000 K$^{*}$ 			& 10,000 K$^{*}$   \\
n$_{e}$ 								& 127 					& 119 						&  489 					& 41 						& 100$^{**}$   \\
W$_{\textrm{abs}}$\,(\AA) 					& 2.0 					& 5.0 						& 0.0 					& 1.0		 				& 1.0$^{av}$ \\
c(H$\beta$) 							& 0.191 					& 0.286 						& 0.044  					& 0.051  					& 0.074    \\
\noalign{\smallskip} \cline{1-6} 
\noalign{\smallskip} \cline{1-6}
\\
\multicolumn{6}{p{0.7\linewidth}}{ ** indicates a default n$_{e}$ of 100 cm$^{-3}$ was assumed.  \te\ values marked with $^{*}$ were not calculated directly, but were assumed to be 10,000 K. $^{av}$ next to the value for W$_{\textrm{abs}}$ indicates that the Balmer line data was not sufficient to derive a value for W$_{\textrm{abs}}$ so an average absorption correction of 1.0 \AA\ was used instead.}

\end{longtable}
\end{singlespace}
\end{landscape}
}
\normalsize

\section{ABUNDANCE DETERMINATION}
\label{sec:abun}

\begin{figure}
\centering
\includegraphics[width=6in]{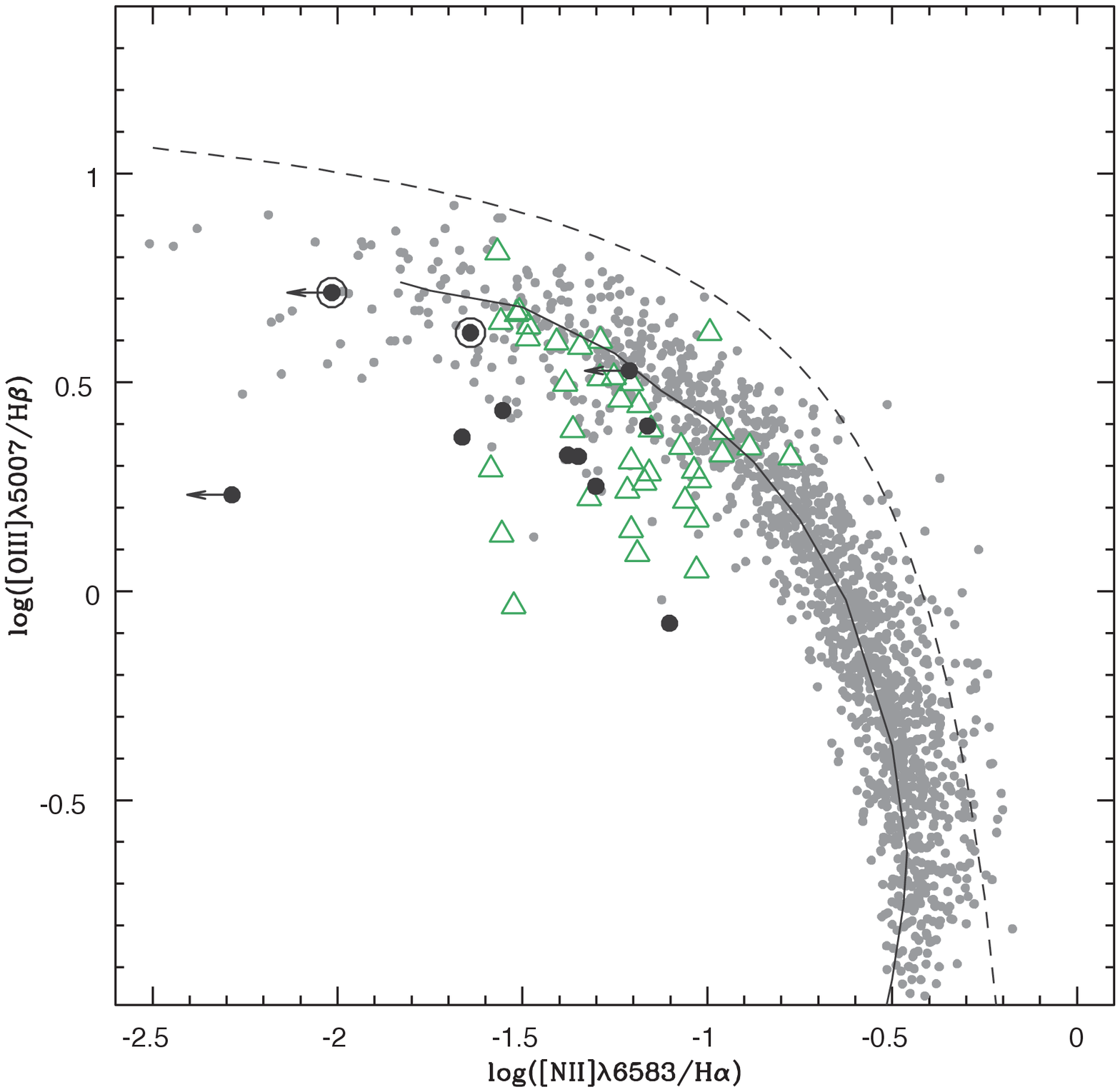}
\caption[Emission-line diagnostic diagram with \hii\ regions from SHIELD galaxies.]{Emission-line diagnostic diagram with each \hii\ region plotted separately.  The black points are from the SHIELD galaxies in this work, those with outer circles are \hii\ regions for which we obtained a measurement of the temperature sensitive \oiiite\ line. The green triangles are galaxies from the ADBS survey \citep{hau13} and the grey points are a large sample of emission line sources from the KISS survey \citep[e.g.,][]{salz00, ms02}.  The solid line traces the star-forming galaxy models of \citet{de86}; the dashed line \citep{kau03} delineates the region of the diagram usually occupied by AGN.}
\label{fig:c4diag}
\end{figure}

The emission line ratio of \nha\ and \ohb\ can be used to plot each \hii-region on a traditional diagnostic diagram \citep[e.g.,][]{BPT, vo87} shown in Figure \ref{fig:c4diag}, which provides a rough diagnostic for excitation and temperature (thus metallicity) of emission nebulae. SHIELD galaxies are plotted on Figure \ref{fig:c4diag} as solid black points; those with circles around them are the two \hii\ regions which displayed a measurable \oiiite\ flux.  The green triangles represent \hii\ regions in low \hi\ mass dIrr galaxies from the Arecibo Dual Beam Survey  (ADBS) which is a similarly selected, but more luminous, sample \citep{hau13}, and the grey points are galaxies from the KPNO International Spectroscopic Survey (KISS) \citep[e.g.,][]{salz00, ms02, salz05c4}. The dashed line differentiates active galactic nuclei (AGN) from star-forming nebulae \citep{kau03} and the solid line represents the locus of high excitation star-forming nebulae from the models of \citet{de86}, which increase smoothly in metallicity from the upper left to the lower right. 

It is clear that \hii\ regions from the SHIELD galaxies lie below and to the left of the majority of the KISS sample and the theoretical high-excitation star-forming model curve.  This is consistent with the manner in which these galaxies were selected and indicates that the observed \hii\ regions in the SHIELD galaxies are generally low-excitation systems.  Emission-line surveys like KISS are much less likely to include these types of low-excitation systems because the key emission lines used for selection in those surveys need to be strong, high equivalent-width lines in order to be detected.  The relatively low-excitation of these systems indicates that either the initial mass function (IMF) was not fully populated or the most massive stars formed in these regions have already evolved off the main sequence. Both possibilities are consistent with the expectation for very low-luminosity galaxies as either would lead to reduced luminosity and surface brightness in comparison to the more well-studied low-metallicity BCDs.  

Three of the \hii\ regions that we observed (AGC 110482\emph{b}, AGC 112521, AGC 111977) had such weak \nii\ lines that we were unable to determine an accurate measurement of \nha.  Thus, these three \hii\ regions have been plotted with arrows indicating the measured \nha\ is an upper limit.  The lack of distinguishable \nii\ emission is consistent with these being among the lowest metallicity regions (see Table \ref{tab:abun}). The \hii\ regions from the SHIELD galaxies all occupy the region of the diagram expected for low-metallicity star-forming systems. However, the \hii\ regions spread over a significant portion of the low-metallicity region while the derived abundances in Table \ref{tab:abun} cover a quite narrow range.  This is consistent with the finding of \citet{hau13} that there is significant scatter in the position of similar metallicity \hii\ regions on the diagnostic diagram.

In order to accurately determine the chemical abundances for emission line nebulae, it is preferable to have a direct measurement of the electron temperature \te.  In the wavelength range covered by our observations, the best way to calculate \te\ is from measuring the \oiiite\ line.  The \oiiite\ and \oiii\ lines are produced by electronic transitions from different energy levels in the O$^{+2}$ ion.  The relative strengths of the lines thus depend on the population of the different energy levels, which is strongly dependent on the temperature of the electrons (and to a lesser extent, the electron density).  \oiiite\ is usually a very weak line that is often difficult to detect.  We were able to detect this line in two of our \hii\ regions, although the detection in AGC 182595 is noisy and thus fairly uncertain.  In order to calculate the nebular abundances for these \hii\ regions we used the ELSA (Emission Line Spectrum Analyzer) program which is described in detail in \citet{elsa}.  

ELSA uses a five-level atom routine with ionization correction factors and a two-region ionization model to calculate abundances.  The five-level atom calculations are based on the work of \citet{hen89} but include multiple updates and improvements \cite[see][]{elsa}. The \oiiite/\oiii\  ratios is used to calculate temperature in the high-ionization region and the [\ion{S}{2}]$\,\lambda \,\lambda6716, 6731$ ratio used to calculate the electron density (in the low-ionization region) using an iterative process similar to that described in \citet{izo06}. The low-ionization region temperature would ideally be calculated from the ratio [\ion{N}{2}]$\lambda$5755/[\ion{N}{2}]$\lambda\lambda$6548,6583, however, [\ion{N}{2}]$\lambda$5755 is a weak line that is usually not detectable in star-forming regions. Therefore, the temperature of the low-ionization region is estimated using $T_{e,O^{+2}}$ and following the method of \citet{pag92}. The ionization correction factors are determined from lines with similar ionization potentials and then used to account for the unseen ionization states. Many uncertainties, including those in line fluxes, reddening correction, plasma diagnostics, and ionic abundances are carried through the ELSA program and propagated properly into the final uncertainties in the abundances.


In cases where \te\ is not directly measurable, strong-line calibrations must be used.  These methods rely on ratios of strong emission lines to estimate an electron temperature and thus metallicity.  Specifically, most strong-line abundance methods are calibrated to a specific oxygen abundance; we will use oxygen abundance to be synonymous with metallicity for the remainder of this paper.  While these methods are very useful, some caution must be employed as they are reliant on both theoretical models and empirical calibration to arrive at an abundance.  

\begin{figure}
\centering
\includegraphics[width=6in]{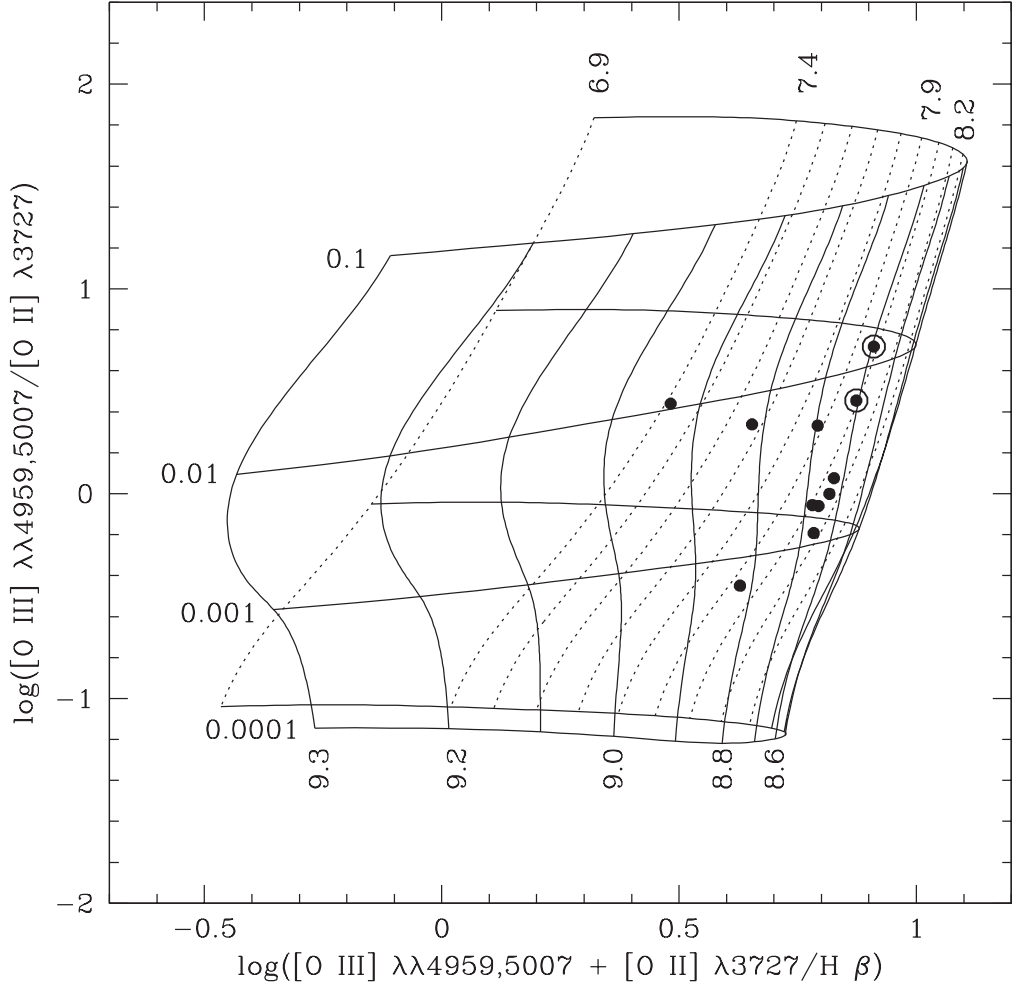}
\caption[McGaugh grids with \hii\ regions from the SHIELD galaxies]{Theoretical grids for estimating abundance from \citet{mcg} plotted with \hii\ regions from the SHIELD galaxies.  The roughly horizontal lines indicate different ionization parameters and the roughly vertical lines represent lines of constant oxygen abundance (expressed as log(O/H)+12).  A line of constant oxygen abundance is displayed for each 0.1 dex increase from log(O/H)+12 = 7.4 to 9.3.  All the objects in this study appear to be located on the ``far-side," or low-metallicity branch (log(O/H)+12 $<$ 8.2), of the surface traced by the models. The points plotted within circles represent those \hii\ regions where the \oiiite\ line was detected.}
\label{fig:c4mcg}
\end{figure}

One of the most robust strong-line methods for determining oxygen abundance is that of McGaugh (1991; henceforth the McGaugh method).  The McGaugh method uses several key abundance ratios:
\begin{description}
\item[R$_{23}$]: log(\rtt)
\item[O$_{23}$]: log(\ott)
\item[[\ion{N}{2}\textrm{]}/[\ion{O}{2}\textrm{]} ]: log(\niioii).
\end{description} 
Each measured \hii\ region is placed on a grid of model emission nebula using the R$_{23}$ and O$_{23}$ parameters as shown in Figure \ref{fig:c4mcg}.  Since the models are degenerate (i.e. there are two possible oxygen abundances that can produce the same values for R$_{23}$ and O$_{23}$) the value of [\ion{N}{2}\textrm{]}/[\ion{O}{2}\textrm{]} is used to break the degeneracy between high- and low- metallicity ``branches."  \hii\ regions are placed on the low-metallicity branch if  [\ion{N}{2}]/[\ion{O}{2}]$< -1.0$ and the high-metallicity branch if [\ion{N}{2}]/[\ion{O}{2}] $> -0.9$. If $-1.0 <$ [\ion{N}{2}]/[\ion{O}{2}] $< -0.9$ the \hii\ regions are deemed to be in the ``turn-around region" where precise determination of the abundance can be tricky.  However, all the \hii\ regions in this work had very low \niioii\ placing them securely onto the low-metallicity branch.  

The degeneracy of the McGaugh models can be troublesome and lead to significant uncertainty when the branch is not well determined.  However, the method is ideal for use with a sample like ours where the branch is unambiguous.  The McGaugh method is a particularly robust strong-line calibration because it uses the total flux in the strong oxygen lines (the R$_{23}$ parameter) to determine oxygen abundance, and also accounts for different hardness of the ionizing spectra by incorporating the O$_{23}$ parameter which greatly improves its accuracy.  The precise value of the oxygen abundance derived from the McGaugh method depends not only on the values of R$_{23}$ and O$_{23}$ but also on the specifics of which models were used to create the grids.  We have used the models described in \citet{mcg} where the highest mass star produced from the IMF is 60 $M_{\odot}$. The exact value of the oxygen abundance was determined by using the position of each \hii\ region and interpolating between the model points to estimate abundance more precisely. 

A final table of abundances as well as key line ratios is presented in Table \ref{tab:abun}.  Column 1 lists the galaxy and specific \hii-region where applicable.  Columns 2$\--$4 contain the line ratios used to calculate the abundance via the McGaugh method.  Column 5 gives the oxygen abundance determined via the McGaugh method (in terms of log(O/H)+12) while Column 6 gives the oxygen abundance determined via the \te\ method when available.  Column 7 lists the error-weighted mean oxygen abundance from the individual \hii\ regions in AGC 110482 and AGC 749237.  The abundance listed in bold is the assumed abundance for each galaxy.  We do not give explicit errors on the McGaugh abundances, but will assume an error of 0.10 dex.

\afterpage{
\begin{landscape}
\begin{table}[c]
\footnotesize
\begin{longtable}[pc]{l c c c c c c c}
\caption{Derived SHIELD Galaxy Abundances} \label{tab:abun} \\

\hline \hline \noalign{\smallskip}
Galaxy		&    log(R$_{23}$)	& log(O$_{23}$) & log([NII]/[OII])	& log(O/H)+12        & log(O/H)+12             &       log(O/H)+12             \\       
			&   	 			& 		 	   &  			& [McGaugh]	     & [$T_{e}$]	       	       &	[w. mean]	      	\\
\noalign{\smallskip} \hline \noalign{\smallskip}
AGC 112521	&  0.482 	 	& \ 0.440   	 &  $<$ -1.732    	& \textbf{7.41}	     		& $\cdots$	     			&	 $\cdots$			\\
AGC 110482	& \multicolumn{5}{c}{\dotfill}												&	 7.81$\,\pm\,$0.09	\\
\hfill \emph{a}   	&  0.827 	 	& \ 0.076   	 &  \ \ \ -1.583    		& 8.05		     		& $\cdots$	     			&	 $\cdots$			\\
\hfill \emph{b}	&  0.910 	 	& \ 0.718   	 &  $<$ -1.685    	& 7.98		     & \textbf{7.74$\,\pm\,$0.08}  		&	 $\cdots$			\\
AGC 111946	&  0.629 	 	& -0.450   	 &  \ \ \ -1.142    		& \textbf{7.94}		     & $\cdots$	     			&	 $\cdots$			\\
AGC 111977	&  0.792 	 	& \ 0.333   	 &  $<$ -1.048    	& \textbf{7.88}		     & $\cdots$	     			&	 $\cdots$			\\
AGC 111164	&  0.654 	 	& \ 0.338   	 &  \ \ \ -1.359    		& \textbf{7.67}		     & $\cdots$	     			&	 $\cdots$			\\
AGC 182595	&  0.874 	 	& \ 0.455   	 &  \ \ \ -1.482    		& 7.99		     	& \textbf{7.75$\,\pm\,$0.13}  		&	 $\cdots$			\\		 
AGC 731457	&  0.784 	 	& -0.193   	 &  \ \ \ -1.413    		& \textbf{8.08}		     & $\cdots$	     			&	 $\cdots$			\\
AGC 749237	&   \multicolumn{5}{c}{\dotfill}												&	 \textbf{8.03$\,\pm\,$0.09}	\\
\hfill \emph{a}   	&  0.817 	 	& -5.30e-4  &  \ \ \ -1.222   		& 8.06		     & $\cdots$	     			&	 $\cdots$			\\
\hfill \emph{b}		&  0.794 	 	& -0.060   	 & \ \ \ -1.415    	& 8.03		     & $\cdots$	     			&	 $\cdots$			\\
\hfill \emph{c}		&  0.781 	 	& -0.055   	 &  \ \ \ -1.428    	& 8.01		     & $\cdots$	     			&	 $\cdots$			\\
\noalign{\smallskip} \hline \hline
\noalign{\smallskip} 
\multicolumn{7}{p{0.73\linewidth}}{Calculated abundances for SHIELD galaxies.  When more than one HII region was present an error weighted mean, listed in Column 7, was calculated.  The abundances listed in bold are the abundances we assumed for each galaxy. 
The $<$ indicates the [NII] line was not measurable so the [NII]/[OII] in this table is only an upper limit. }
\end{longtable}
\end{table}
\end{landscape}
\clearpage
}
\normalsize


The abundances derived via the McGaugh and \te\ methods are not in perfect agreement with each other which is consistent with what has been reported by other authors. The inconsistency is often considered an effect of intrinsic errors in the strong-line calibration methods since they rely on indirect determinations of \te.  While the \te\ method provides a direct measurement of the electron temperature it is also subject to some level of uncertainty and bias as it has been shown to favor higher temperature regions in the case that there are temperature fluctuations in the nebula being observed \citep{pei67}.   Thus, it can be difficult to determine which method more accurately yields the ``correct" abundance.

 Some authors \citep[e.g.,][]{kew08, zah12} have suggested that empirical calibrations should be made to correct for systematic differences between various abundance determination methods.  With only two \te\ abundances, we are unable to gauge whether we see a systematic offset in our data.  However, if we use the sample of low-luminosity Local Volume Legacy (LVL) galaxies analyzed in \citet{berg12}, which we will use as a comparison set in Section \ref{sec:analysis}, we can determine a reasonable empirical correction and use it to reconcile the two abundance methods.  We have assumed the \te\ abundances listed in \citet{berg12} then calculated McGaugh abundances from the fluxes given in that paper, using the same method and models used for our galaxies.   When the two different abundance calculation methods are compared we can see that there does appear to be a systematic offset.  

A plot of abundance difference (calculated as the McGaugh abundance minus the \te\ abundance)  vs. \te\ abundance for the \citet{berg12} sample is shown in Figure \ref{fig:c4mcg_te}.  The red diamonds are \hii\ regions from the the LVL galaxies in \citet{berg12} and the black points are the \hii\ regions  with \te\ abundances from the SHIELD galaxies. Since it is unclear whether this difference is a constant offset or if it varies with metallicity we have only used \hii\ regions within the range of metallicities that is consistent with our sample (7.4 $\leq$ log(O/H)+12 $\leq$ 8.1) and exclude the obvious outlier.  The points from the \citet{berg12} sample which were used to calculate the mean offset are plotted as filled red diamonds and those that were excluded are plotted as open red diamonds.  The mean offset that we calculated was $0. 076$ dex, indicating that on average the McGaugh abundances were $0.076$ dex higher than those calculated using the \te\ method 
  This offset will later be used as a ``correction" when comparing our sample to others from the literature.

\afterpage{
\begin{figure}
\centering
\includegraphics[width=6in]{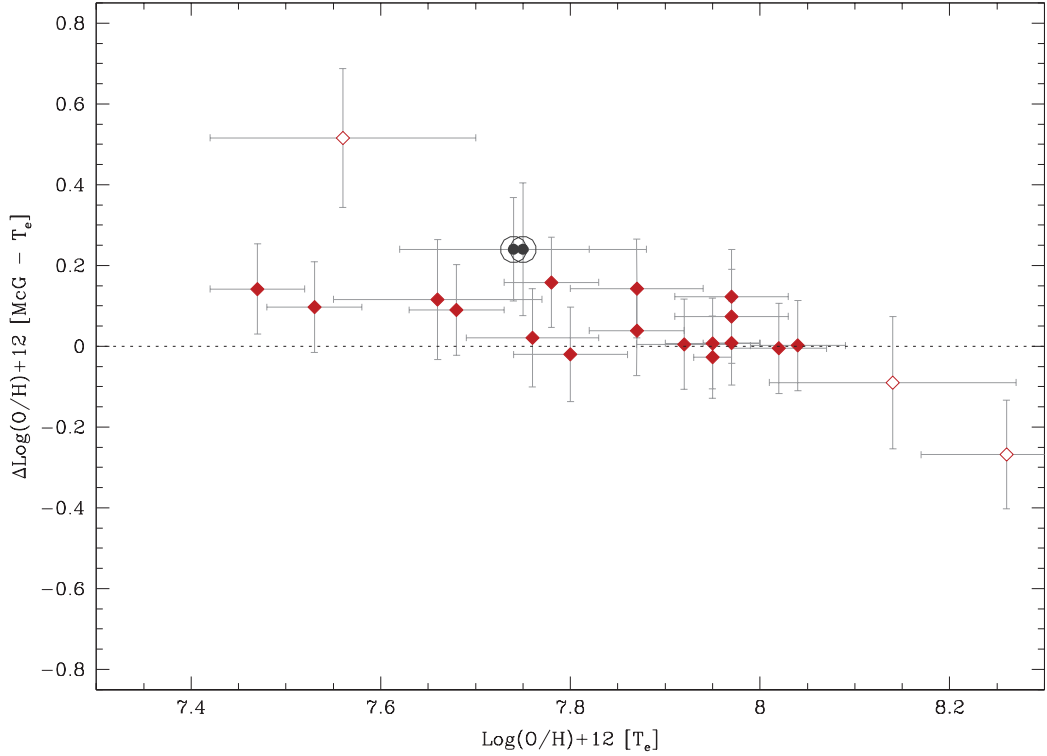}
\caption[Comparison of derived McGaugh and \te\ abundances.]{Comparison of derived McGaugh and \te\ abundances.  Black points within circles are \hii\ regions from galaxies in the SHIELD sample with \te\ derived abundances.  Red diamonds represent galaxies from \citet{berg12}; filled diamonds are those used to calibrate the empirical offset between the two abundance methods, open diamonds are either obvious outliers or outside of the abundance range that we considered for the offset calculation.}
\label{fig:c4mcg_te}
\end{figure}
}

``Corrected" average abundances for the SHIELD galaxies are compiled in Table \ref{tab:abuncor}.  When more than one \hii\ region was measured a weighted mean of the multiple \hii\ region abundances was used.  Notice that the abundances for AGC 112521 and AGC 111164 qualify them both as XMD galaxies which is consistent with them being the two lowest luminosity galaxies for which we obtained usable spectra.

\afterpage{
\begin{table}[c]
\begin{longtable}[pc]{l c l}
\caption{``Corrected" SHIELD Galaxy Abundances} \label{tab:abuncor} \\

\hline \hline \noalign{\smallskip}
Galaxy		&    log(O/H)+12   & Note   \\       
\noalign{\smallskip} \hline \noalign{\smallskip}
AGC 112521	& 7.33$\,\pm\,$0.10  &    corrected McGaugh, \textbf{XMD galaxy}\\
AGC 110482	& 7.79$\,\pm\,$0.07  &	weighted mean of \emph{a} (corrected) and \emph{b} (\te)\\
AGC 111946	& 7.86$\,\pm\,$0.10  &	 corrected McGaugh,		\\
AGC 111977	&  7.80$\,\pm\,$0.10		     & corrected McGaugh		\\
AGC 111164	&  7.59$\,\pm\,$0.10		     & corrected McGaugh, \textbf{XMD galaxy}			\\
AGC 182595	&  7.75$\,\pm\,$0.13  		& \te\ abundance		\\		 
AGC 731457	&  8.00$\,\pm\,$0.10	  &	corrected McGaugh	\\
AGC 749237	&  7.95$\,\pm\,$0.06  & weighted mean of \emph{a, b,} and \emph{c} (all corrected)	\\
\noalign{\smallskip} \hline \hline
\noalign{\smallskip} 
\multicolumn{3}{p{1\linewidth}}{Corrected abundances for SHIELD galaxies (see text for correction).  For AGC 110482 and AGC 749237 a weighted mean of the abundances for each HII regions was used. Both AGC 112521 and AGC 111164 qualify as XMD galaxies.}
\end{longtable}
\end{table}
\clearpage
}
\normalsize


\section{ANALYSIS}
\label{sec:analysis}

There is a well-established relationship between the luminosity and metallicity of galaxies, where lower luminosity corresponds to lower metallicity, which has been shown to extend down to low-luminosity dIrrs \citep[e.g.,][]{Leq79, skill89a, pil01, ms02, trem04, salz05b, lee06c4, vzh06, berg12, hau13, leop_skill}.  The underlying cause of this relationship is not fully understood, but it is believed to trace a more fundamental mass-metallicity (M-Z) relationship, where luminosity is representative of stellar mass.  Studies of the luminosity-metallicity (L-Z) relationship that use near-infrared (NIR) luminosities (which is better correlated with stellar mass) show decreased scatter \citep[e.g.,][]{lee04, salz05b, lee06c4}, implying the M-Z relationship is the more fundamental physical relationship.  The origin of the relationship, however, is still unclear but is a key to understanding the star-formation histories and chemical evolution of dwarf galaxies. 

Low luminosity galaxies (e.g., $M_{B} \gtrsim -13$), which should also be the most metal-poor galaxies, are a relatively understudied population, thus the nature of the L-Z and M-Z relationships in the very low-metallicity regime remains indeterminate. The XMD galaxies which have been studied are mostly BCDs with very high star-formation densities and relatively high star-formation rates and are not believed to be representative of the majority of the very low-mass dIrr population.  A much more in-depth analysis of extremely low-metallicity systems, including low surface brightness dIrrs, is necessary to characterize these relationships.  In Figure \ref{fig:lz} we have plotted the SHIELD sample in luminosity-metallicity diagrams using oxygen abundance (plotted as log(O/H)+12) to represent metallicity with $M_{B}$ on the luminosity axis.  On Figure \ref{fig:lz}\textbf{b} and \textbf{c} we have also included comparison galaxies from the literature, described in more detail in the paragraph below. The SHIELD galaxies (black points) within an outer circle represent galaxies where a \te\ abundance was measured. The \te\ abundance in Table \ref{tab:abun} has been adopted for AGC 110482\emph{b} and the weighted mean of the 3 \hii\ regions in AGC 749237 has been adopted for that system.  

Our comparison sample is described in Table \ref{tab:samp}.  It has been broken into two pieces: the ``primary" sample (plotted as filled symbols) is composed of only those galaxies with velocity-independent distances \emph{and} \te\ abundances from \citet{lee06c4, vzh06}, and \citet{berg12}. The ``expanded" sample (plotted as open symbols) includes galaxies from \citet{berg12} and \citet{vzh06} with velocity-based distances or strong-line abundances as well as the ADBS sample from \citet{hau13}.  The ADBS sample is of particular interest as it was chosen in a similar manner to the SHIELD sample (e.g., ADBS is a blind \hi\ survey) and thus has similar selection biases.  The ADBS galaxies have velocity-derived distances and the abundances were derived using the same techniques presented in this work with a mix of both the \te\ and McGaugh methods used to derive the abundances (see Haurberg et al.~2013 for details).

\afterpage{
\begin{table}[c]
\begin{longtable}[pc]{l l l }
\caption{Comparison Sample} \label{tab:samp} \\

\hline \hline \noalign{\smallskip}
Sample		&     Reference  		& Symbol     \\       
\noalign{\smallskip} \hline \noalign{\smallskip}
\textbf{Primary}		&  \multicolumn{2}{l}{\textit{Stellar based distances and \te\ abundances}}        \\
				& 	 \citet{berg12}		&  Filled Red Diamonds \\
		  		&  	 \citet{lee06c4}  		&  Filled Blue Points	     \\
				&	\citet{vzh06}		&  Filled Orange Squares \\
\noalign{\smallskip} 			
\textbf{Expanded}		& \multicolumn{2}{l}{\textit{Velocity determined dist. and/or strong-line abundance}}	\\
			&	 \citet{hau13}	     	& Open Green Triangles \\
			&  	  \citet{berg12}	&  Open Red Diamonds \\
			&	  \citet{vzh06}	& Open Orange Squares \\
\noalign{\smallskip} \hline \hline
\noalign{\smallskip} 
\multicolumn{3}{c}{}
\end{longtable}
\end{table}
\normalsize
\clearpage
}

Figure \ref{fig:lz}\textbf{a} shows just the SHIELD galaxies, while Figure \ref{fig:lz}\textbf{b} and \textbf{c} include comparison samples from the literature.  Figure \ref{fig:lz}\textbf{b} includes the primary and expanded comparison samples while Figure \ref{fig:lz}\textbf{c} shows only the primary comparison sample and the SHIELD galaxies.  As indicated, the abundances from the SHIELD galaxies in Figure \ref{fig:lz}\textbf{c} have been corrected, meaning the McGaugh abundances have been adjusted by the offset calculated in Section \ref{sec:abun} in order to be consistent with the \te\ scale. 

\afterpage{
\begin{figure}
\centering
\includegraphics[width=5.5in]{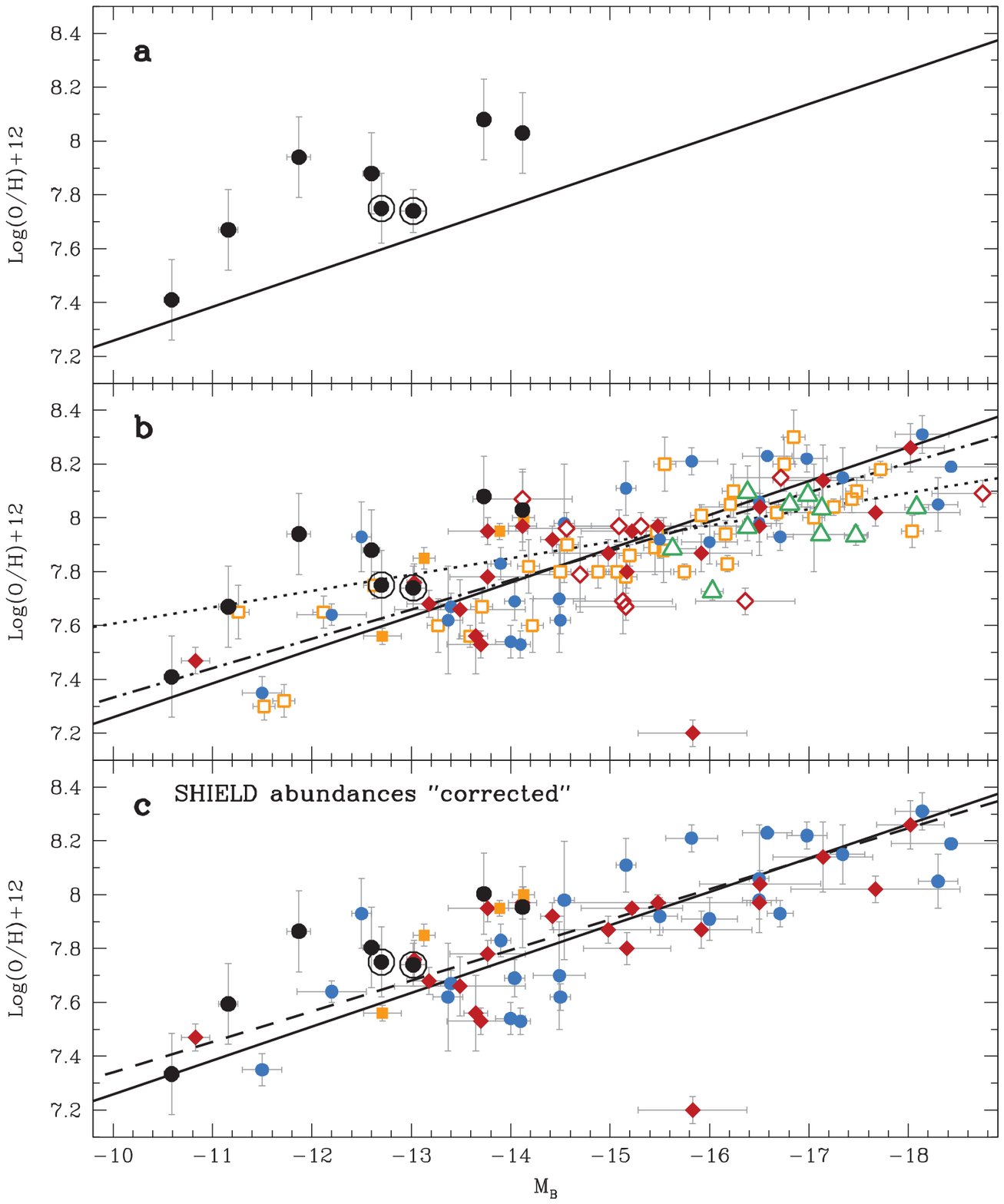}
\caption[Luminosity-metallicity diagrams displaying the SHIELD galaxies and a comparison sample from the literature.]{Luminosity-metallicity diagrams. The SHIELD sample is displayed with filled black points; dwarf irregular galaxies from the literature are shown with various symbols, filled points represent the ``primary'' comparison sample and open points the ``expanded'' comparison sample (see Table \ref{tab:samp}):  \citet{lee06c4} = blue points, \citet{hau13} = green triangles, \citet{vzh06} = orange squares, and \citet{berg12} = red diamonds. Figure \ref{fig:lz}\textbf{b} shows the entire comparison sample and the uncorrected SHIELD abundances (Table \ref{tab:abun}); \textbf{c} includes galaxies from only the primary comparison sample and the ``corrected" SHIELD abundances (Table \ref{tab:abuncor}).  The lines represent fits described in the text and Table \ref{tab:fits}.}
\label{fig:lz}
\end{figure}
}

The data shown in Figure \ref{fig:lz} were fit using a bivariate linear regression method weighting the errors in both x and y. The method we used is based on that outlined in \citet{akber96} and references therein.  The fit to the primary comparison sample is shown on all three panels in Figure \ref{fig:lz} as a solid line. The dash-dot line shown in Figure \ref{fig:lz} \textbf{b} is the fit to all of the comparison sample plus the SHIELD galaxies and the dashed line on Figure \ref{fig:lz}\textbf{c} is the fit to the primary comparison sample plus the SHIELD galaxies (using the corrected abundances). The XMD outlier, UGC 5340 ($M_{B} = -15.83$, $\log$(O/H)$+12 = 7.20$), was excluded from all of the fits. The fit that is shown with a dotted line on Figure \ref{fig:lz}\textbf{b} was obtained by fitting only the \hi\,-\,selected galaxies (SHIELD and ADBS samples) and is interesting since it is notably different from the other fits.  The fit parameters and samples which were used to obtain them are described in Table \ref{tab:fits}.  The reduced $\chi^{2}$ values listed in that table are calculated for each fit and compared to the sample plotted in the indicated figure; $\chi^{2}_{\textrm{red}}$ was calculated assuming abundance to be the dependent variable. We chose not to include a fit to the SHIELD sample alone because it is a statistically small sample that covers a limited luminosity range, thus it produced significant uncertainties in the quality of the fit.
 
\afterpage{
\begin{table}[c]
\scriptsize
\begin{longtable}[c]{l c c c c c c}
\caption{L-Z and M-Z Relationship Fit Parameters} \label{tab:fits} \\

\hline \hline \noalign{\smallskip}
Sample					&    	Slope 		& Intercept 			& Symbol		 &   \multicolumn{3}{c}{Goodness of Fit ($\chi^{2}_{\textrm{red}}$)}\\
\noalign{\smallskip} \hline \noalign{\smallskip}
\multicolumn{4}{l}{\textbf{log(O/H)+12 vs.~$M_{B}$ $\dotfill$}} & \textbf{\ref{fig:lz}a} & \textbf{\ref{fig:lz}b} & \textbf{\ref{fig:lz}c} \\
\noalign{\smallskip}
Primary Comparison					&  -0.125$\,\pm\,$0.008 	 	&  6.01$\,\pm\,$0.12   	 &  Solid     		& 8.54 & 13.23   & 12.42	\\
SHIELD+ Primary + Expanded 			& -0.110$\,\pm\,$0.006	 	&  6.24$\,\pm\,$0.10   	 &  Dot-Dash    		&  $\cdots$& 11.43   & $\cdots$   \\ 
HI-Selected (SHIELD + ADBS)			&  -0.061$\,\pm\,$0.012 		&  7.00$\,\pm\,$0.24    	 &  Dotted	 		& $\cdots$ & 12.91  & $\cdots$    \\ 
SHIELD$^{\dag}$ + Primary			&  -0.113$\,\pm\,$0.007 	 	&  6.21$\,\pm\,$0.11   	 &  Dashed   	       & $\cdots$ & $\cdots$ & 11.52	 \\ 
\noalign{\smallskip}
\cline{1-7}
\noalign{\smallskip}\noalign{\smallskip}
\multicolumn{6}{l}{\textbf{log(O/H)+12 vs.~$M_{4.5}$ $\dotfill$}} & \multicolumn{1}{c}{\textbf{\ref{fig:mz}a}}\\
\noalign{\smallskip}
Primary Comparison					&  -0.112$\,\pm\,$0.007 	 	&  5.94$\,\pm\,$0.13   	 &  Solid     	& &	& \multicolumn{1}{c}{10.07}\\
SHIELD$^{\dag}$ + Primary			&  -0.110$\,\pm\,$0.008 	 	&  5.98$\,\pm\,$0.14   	 &  Dashed   	& &   & \multicolumn{1}{c}{9.76}\\
\noalign{\smallskip}
\cline{1-7}
\noalign{\smallskip}\noalign{\smallskip}
\multicolumn{6}{l}{\textbf{log(O/H)+12 vs.~log($M_{\star}$}) $\dotfill$} & \multicolumn{1}{c}{\textbf{\ref{fig:mz}b}}\\
\noalign{\smallskip}
Primary Comparison					&  0.29$\,\pm\,$0.09 	 	&  5.64$\,\pm\,$0.71   	 &  Solid    		& & 	&  \multicolumn{1}{c}{11.42}\\
SHIELD$^{\dag}$ + Primary			&  0.30$\,\pm\,$0.25 	 	&  5.60$\,\pm\,$1.86   	 &  Dashed  	 	& & & \multicolumn{1}{c}{11.57}\\
\noalign{\smallskip} \hline \hline
\noalign{\smallskip} 
\multicolumn{7}{p{0.9\textwidth}}{The results of various fits to the L-Z and M-Z relationship.  $^{\dag}$ indicates that corrected McGaugh abundances were used for the SHIELD galaxies. The ``goodness of fit" parameter is the reduced $\chi^{2}$ value for the given fit when compared with the data plotted on the indicated panel or figure.  See text for details of the fitting process and different samples.}
\end{longtable}
\end{table}
\clearpage
}
\normalsize


The various fits are in general agreement within the errors (with the exception of the ``\hi\,-\,sample" fit which is discussed in the following paragraph). In all cases, the inclusion of the SHIELD sample results in a more shallow slope, as could easily be predicted from inspection of Figure \ref{fig:lz}. The SHIELD galaxies clearly lie above literature galaxies of similar luminosity even when the corrected abundances are used.  This is certainly an intriguing trend, but since we have a relatively small sample it is possible that this is due to intrinsic scatter in the relationship owing to differing mass-to-light ratios in the $B$-band or uncertainty in measurement of both luminosity and abundance.  However, it may represent something more significant such as a fundamental difference in the samples or a flattening of the L-Z relationship at low-luminosity.  

The fit obtained from the \hi\,-\,selected samples (dotted line) is interesting as it is significantly different from the other fits.  It has a slope that is substantially shallower than the average slope of the other fits, but provides the best fit to the SHIELD sample according the reduced $\chi^{2}$ test. This indicates that the sample selection process may be a contributing factor and may suggest underlying physical differences between samples selected from \hi\ surveys and those selected from optical catalogs or emission line-surveys.  For consistency, we used the uncorrected abundances for Figure \ref{fig:lz}\textbf{b} and when fitting this \hi\ selected sample since the ADBS galaxies have a mix of strong-line and \te\ derived abundances.  When the corrected abundances are used (Fig. \ref{fig:lz}\textbf{c}) the obvious discrepancy between the SHIELD sample and the fit from the comparison sample is reduced.  However, even when the corrected abundances are used, all but one of the SHIELD galaxies lie above the fit line indicating some level of systematic difference that is unlikely to be due solely to intrinsic scatter in the relationship.  These issues are further discussed in Section \ref{sec:disc}. 

\begin{figure}
\centering
\includegraphics[width=6in]{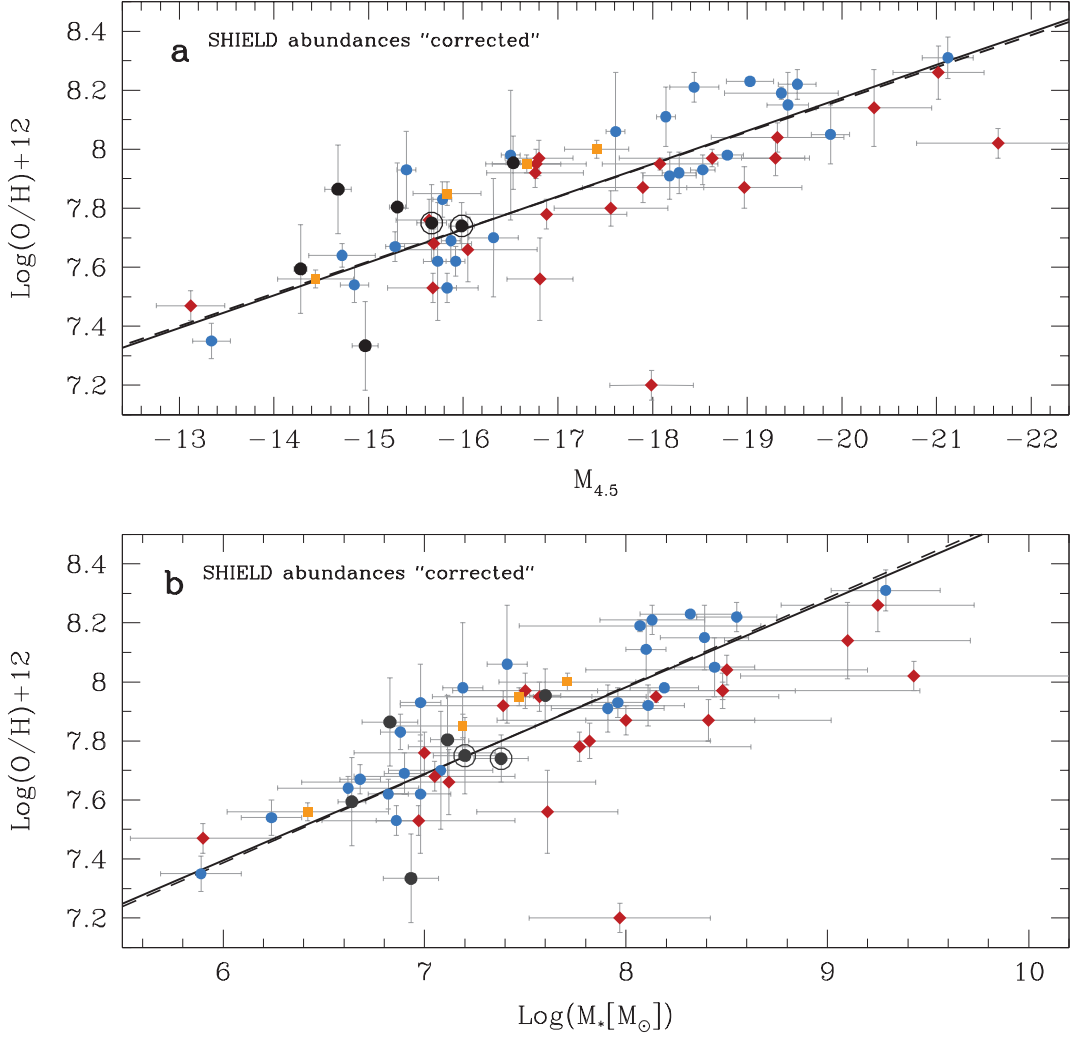}
\caption [NIR L-Z and M-Z relationships for SHIELD galaxies and comparison sample.] {NIR-L-Z relationship and M-Z relationship.  Symbols and colors are the same as Figure \ref{fig:lz}.  The two fits shown are the fit to the primary comparison sample only (solid line) and fits to the sample including the SHIELD galaxies (dashed line).}
\label{fig:mz}
\end{figure}

Figure \ref{fig:mz}\textbf{a} shows another version of the L-Z diagram.  This version uses the corrected abundances for the SHIELD galaxies, like Figure \ref{fig:lz}\textbf{c}, but uses the NIR luminosity (4.5$\micron$ band; $M_{4.5}$) instead of the $B$-band luminosity as the luminosity metric.  Since we do not have reliable 4.5$\micron$ luminosities for one of the SHIELD galaxies for which we measured an abundance (AGC 731457), this version of the L-Z relationship only features 7 of the galaxies in the SHIELD sample.  The SHIELD galaxies are more coincident with the comparison data in the NIR version of the L-Z diagram than they are in Figure \ref{fig:lz}.  The fits derived from fitting the comparison sample (solid line) are nearly identical to those calculated when the SHIELD data are added to the set (dashed line). Despite the similarity of the fits, the majority of the SHIELD galaxies still appear above the fit line on this diagram, again suggesting the possible presence of a systematic difference from the comparison sample.  The fit parameters for Figure \ref{fig:mz} are given in Table \ref{tab:fits}. 

In Figure \ref{fig:mz}\textbf{b} the mass metallicity (M-Z) relationship is shown. For the comparison sample, stellar mass was derived using the 4.5$\micron$ flux and the $m_{4.5}- K$ ($K$-band magnitude) color following \citet{lee06c4}. The SHIELD IR observations do not include the K-band, so stellar masses for the SHIELD sample had to be derived in a different manner.  We used the 3.6 and 4.5$\micron$ magnitudes and followed the method of \citet{esk12}.  Since there can be significant uncertainties associated with stellar mass calculations, we performed an independent derivation to check against the mass obtained with the method of \citet{esk12}.  As a check we used a straight-forward conversion of 3.6$\micron$ luminosity into mass assuming a mass-to-light ratio of 0.5 \citep{mcg13} and $M_{\odot, 3.6\micron} = 3.24$ \citep{oh08}.  The difference between the mass derived using \citet{esk12} and that derived from our ``straight-forward" conversion of 3.6\micron\ luminosity was very small; the average difference was only 0.008 dex.  While we can not confirm that our mass estimates are fully consistent with the \citet{lee06c4} method used by \citet{berg12}, we do feel confident that the stellar-mass estimates from \citet{esk12} for the SHIELD galaxies given in this paper are consistent with other IR-based mass derivation methods. Due to the use of different stellar-mass calculation methods and the inherently difficult nature of determining stellar masses, we proceed with some caution concerning the results derived using these mass estimates.

Similar to the NIR-LZ relationship, the SHIELD galaxies appear consistent with the literature samples in the M-Z relationship shown in Figure \ref{fig:mz}\textbf{b} and the inclusion of the SHIELD data has very little effect on the results of the fit.  The fit parameters and $\chi^{2}_{\textrm{red}}$ for the M-Z relationship are included in Table \ref{tab:fits}.

\section{DISCUSSION}
\label{sec:disc}

Various physical mechanisms have been suggested to explain the origin of the M-Z relationship for dIrrs, but the ejection of enriched gas from low-mass galaxies via supernova and stellar winds is one of the most commonly invoked \citep[e.g.,][]{trem04}.   In this scenario, the relationship arises as an effect of an enriched-gas-retention sequence where more massive galaxies are better able to retain enriched ejecta thus producing higher gas phase metallicities.  While the retention of metal-enriched gas is likely a contributing factor, most observational studies that indicate the presence of significant gas outflows from isolated dwarf galaxies are based primarily on BCDs \citep[e.g.,][]{pap94, mar95, mar05, oey08} and may not apply to dIrrs with much less active modes of star-formation. \citet{stin07}, among others, have proposed models for the evolution of dwarf galaxies where star-formation occurs in patterns of relatively strong starbursts that are quenched by supernova feedback and then followed by periods of more moderate star formation (the so called ``breathing" model).  This ``bursty" star formation history may supply a mechanism to remove enriched gas while allowing for periods of more quiescent evolution.  However, it is unclear if the majority of dIrrs ever undergo the massive starbursts that are necessary to drive such a model as photometric studies of BCDs suggest that the structural properties of these systems are fundamentally different than more ``normal" low-surface brightness dIrrs \citep{pap96, dou99, mar99, sn99, janslaz}. Thus other evolutionary scenarios may need to also be considered to explain the observed trends at the low-mass (and low-luminosity) end of the L-Z and M-Z relationships.

As an alternative to bursty star-formation histories for isolated dIrrs, \citet{gav13} suggest that many observational features of isolated dIrrs can be explained \emph{without} invoking winds or large scale outflows, but instead rely on continuous infall or cooling of primordial gas and continuous star formation throughout the lifetime of the galaxy.  This model provides interesting implications in the context of the results presented in this paper.  

In the scenario proposed by \citet{gav13} the galaxies are modeled as a single region and the star formation rate is controlled only by the amount of gas available and an assumed star formation efficiency. This model does not allow for ``bursts" of rapid star formation but instead produces a smooth, continuous (though not constant) star-formation history for the galaxy.  
Their results suggest that with moderate to low star-formation efficiencies this continuous type of evolution can plausibly reproduce the observed dIrr trends and the luminosity-metallicity relationship. The \citet{gav13} models imply that the observed dispersion in abundance seen at a given $B$-band luminosity is a result of differing star-formation histories, owing to different star-formation efficiencies and/or different gas-collapse timescales. This is intriguing in the context of the work we have presented here, because it suggests the offset of the SHIELD sample seen in the $B$-band L-Z relationship may be indicative of different global star-formation history compared to optically selected samples.  Since the SHIELD sample has specifically been chosen to represent some of the lowest-luminosity and lowest-surface brightness galaxies, it is not unreasonable to assume that our selection may be biased to select galaxies with different star-formation histories than those selected from optical catalogs.  

\citet{tas08} produced similar results to \citet{gav13} suggesting that galaxian outflow winds are not necessary to reproduce the observed L-Z trends and attribute the inefficient star-formation in low-mass systems to inefficient gas cooling (i.e. longer collapse timescales) because of the more extended neutral gas distributions often observed in lower mass systems.  Since there are only a limited number of low-mass dIrr samples where high quality optical and \hi\ data currently exist, it is very difficult to analyze whether this scenario is observationally consistent.  However, we suggest this is an idea worthy of further investigation. Further studies with the SHIELD project, and similar projects focusing on low-mass ALFALFA sources, should lead to a better understanding of the nature of the gas distributions in low-mass systems and the role that gas dynamics play in their chemical evolution.

The SHIELD galaxies all seem to be ``offset" above the fit line from the comparison sample in the $B$-band L-Z relationship.  When the ``corrected" abundances are used, the offset, with respect to the fit from the comparison sample, is reduced, yet all but one of the SHIELD galaxies still appears above the fit line.  This offset becomes even less pronounced when $M_{B}$ is replaced with $M_{4.5}$ and disappears when stellar mass ($M_{\star}$) is used instead. This indicates that, on average, the mass-to-($B$-band) light ratio for the SHIELD galaxies differs from the comparison sample.  This is plausibly consistent with our previous suggestion, that our selection criteria may have biased our sample toward galaxies with a certain star-formation history. However, the general trend simply suggests that the SHIELD galaxies are \emph{too massive} for their $B$-band luminosity.    
This may arise because at least some portion of the SHIELD galaxies do not approach the quasi-continuous star-formation history that has been suggested for more massive dIrrs by \citet{vz01}.  A stochastic star-formation history that is similar to that suggested in \citet{vz01} but not as continuous (owing to the lower mass of the system) could cause the $B$-band luminosity to be less reflective of the current stellar mass in these galaxies.


The results presented in this paper do not necessarily clarify any of the issues associated with the chemical evolution of dIrrs, but they do suggest that varied star-formation histories need to be considered along with the effects of enriched gas outflows in order to adequately explain the observed trends.  Additionally, our data suggest that while the M-Z relationship may be more fundamental, understanding the relationship of the mass-to-light ratio could reveal details of the star-formation and chemical evolution in these systems.  This sample is just a first step in beginning to fill in the very low-luminosity range of dwarf irregulars.

\section{CONCLUSIONS}
\label{sec:concl}

We have used long-slit spectra from the Mayall 4m telescope at KPNO to determine the oxygen abundances for 8 of the 12 SHIELD galaxies. For two galaxies we were able to measure the temperature sensitive \oiiite\ line and thus directly determine the electron temperature for at least one \hii\ region.  We calculated abundances for the other 6 galaxies using the strong-line method of \citet{mcg}.  We calculated an empirical correction to the McGaugh abundance scale using the dIrr sample from \citet{berg12}, along with the two \hii\ regions for which we detected \oiiite, allowing us to put our McGaugh abundances on a consistent scale with those that have a directly determined \te.  The galaxies in our sample have  well-constrained distances (HST; McQuinn et al. 2014), optical luminosities (WIYN 3.5m BVR; N.~Haurberg et al.~in preparation), and NIR luminosities (Spitzer 3.6 and 4.5$\micron$; Cannon, Marshall et al.~in preparation). We compared the L-Z and M-Z relationships derived from our results with a substantial comparison set from the literature that also have well-determined abundances, distances, and optical and NIR luminosities \citep[][and references therein]{lee06c4, vzh06, berg12}. We additionally included dIrr galaxies with well-determined abundances that were selected from the ADBS blind \hi\,-\,survey \citep{hau13} in our comparison sample.  While more luminous than our data set, the latter sample is particularly interesting because the method of selection was very similar to that used for the SHIELD data set. 

When the L-Z relationship was examined using the $B$-band luminosity the SHIELD galaxies appear offset from the comparison sample, but that offset disappears when derived stellar mass is used instead.  We suggest that this may indicate that a range of star-formation histories and chemical evolution scenarios for dwarf irregulars should be considered. The possible effects that such scenarios may have on the the luminosity-metallicity and mass-metallicity relationships warrant further examination and exploration. While the M-Z relationship may be more well-constrained, the L-Z relationship may prove observationally important for disentangling the origins of that relationship. A larger and more comprehensive sample is needed and a more exhaustive comparison between an array of models and multiple observational parameters should be done before any model is too heavily favored. Such a comparison is beyond the scope of this work.

In this work, we identified two new XMD galaxies in the local universe, AGC 112521 and AGC 111164, and showed that blind \hi\,-\,surveys like ALFALFA are effective in providing candidate XMD dIrrs \citep[e.g.,][]{leop_skill}. As follow-up studies of the low-\hi\ mass sources from the ALFALFA survey continue, a larger sample of low-luminosity gas-rich dIrrs will become available and allow us to start to unravel some of the puzzles concerning the chemical evolution of dwarf galaxies.  Additionally, current work on resolved stellar populations in the SHIELD galaxies and the Leo P dwarf promise to further illuminate the star-formation histories in these very low-mass galaxies and lead to a deeper understanding of galaxian chemical evolution.

\acknowledgements
NCH and JJS acknowledge financial support for this project from Indiana University, including a Dissertation Year Fellowship to NCH from the College of Arts and Sciences. NCH also acknowledges the financial support from Knox College and, in addition, NCH received support from the Indiana Space Grant Consortium in the form of a graduate fellowship. JMC is supported by NSF grant AST-1211683.  We are grateful for  the professional support provided by the staff of Kitt Peak National Observatory during our two observing runs.
\clearpage



\end{document}